\documentclass[journal=mamobx,manuscript=article,layout=twocolumn]{achemso}

\let\oldmaketitle\maketitle
\let\maketitle\relax
\mciteErrorOnUnknownfalse
\SectionNumbersOn

\usepackage{amsmath,bbold,bm,amssymb,mathtools,nccmath}

\usepackage{chemformula}
\usepackage[T1]{fontenc}
\usepackage{soul,cuted,color}
\usepackage{tcolorbox}
\usepackage{xparse}
\usepackage{etoc}

\usepackage{graphicx}
\usepackage{xcolor}
\usepackage{multirow}
\usepackage{colortbl,booktabs}
\usepackage{placeins}
\usepackage{float}
\usepackage{xr-hyper}
\usepackage{uri}
\usepackage[linkbordercolor=green]{hyperref}
\usepackage{xurl}
\usepackage{soul}

\hypersetup{
     colorlinks=true,
     linkcolor=blue,
     filecolor=blue,
     urlcolor=cyan,
     citecolor =blue,
     breaklinks=True
     }

\definecolor{amber}{rgb}{1.0, 0.49, 0.0}
\definecolor{auburn}{rgb}{0.43, 0.21, 0.1}

\makeatletter
\DeclareRobustCommand{\rev}{%
  \ifmmode
    \expandafter\@rev@math
  \else
    \expandafter\@rev@text
  \fi
}
\newcommand{\@rev@math}[1]{%
  \text{\hl{$#1$}}%
}
\newcommand{\@rev@text}[1]{%
  \textcolor{red}{\hl{#1}}%
}
\makeatother

\newcommand{\upperRomannumeral}[1]{\uppercase\expandafter{\romannumeral#1}}
\renewcommand{\thesection}{\Roman{section}}
\renewcommand{\thesubsection}{\thesection.\Roman{subsection}}
\definecolor{ao}{rgb}{0.0, 0.5, 0.0}

\makeatletter

\newcommand*{\addFileDependency}[1]{
\typeout{(#1)}

\@addtofilelist{#1}
\IfFileExists{#1}{}{\typeout{No file #1.}}
}\makeatother

\author{Jude Ann Vishnu}
\email{jvishnu@uni-mainz.de}
 \affiliation{Institute of Physik, Johannes Gutenberg-Universit\"at Mainz,
     Staudingerweg 7-9, 55128 Mainz, Germany}

\author{Torsten Gereon Linder}
\affiliation{Department of chemistry, Johannes Gutenberg-Universit\"at Mainz,
    Duesbergweg 10-14, 55128 Mainz, Germany}

\author{Sebastian Seiffert}
\affiliation{Department of chemistry, Johannes Gutenberg-Universit\"at Mainz,
    Duesbergweg 10-14, 55128 Mainz, Germany}

\author{Friederike Schmid}
\email{friederike.schmid@uni-mainz.de}
 \affiliation{Institute of Physik, Johannes Gutenberg-Universit\"at Mainz,
     Staudingerweg 7-9, 55128 Mainz, Germany}

\title[blah]{Structure and dynamic evolution of interfaces between polymer solutions and gels and polymer interdiffusion: A Molecular dynamics study}


\begin{document}

\begin{tocentry}
\includegraphics[width=8.3cm]{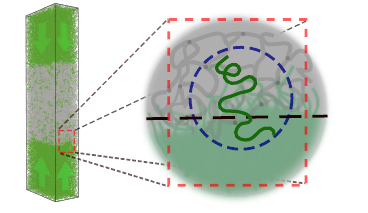}
\end{tocentry}

 \twocolumn[
 \begin{@twocolumnfalse}
 \oldmaketitle

\begin{abstract}
 Letting free polymers diffuse from solution into a crosslinked polymer gel is often a crucial processing step in the synthesis of multiphase polymer-based gels, e.g., core-shell microgels. Here we use coarse-grained molecular dynamics simulations to obtain molecular insights into this process. We consider idealized situations where the gel is modeled as a regular polymer network with the topology of a diamond lattice,  and all free polymers and strands have the same length and consist of the same type of monomer.
After bringing the gel and the polymer solution into contact, two time regimes are observed: An initial compression of the gel caused by the osmotic pressure of the solution, followed by an expansion due to swelling. We characterize the time evolution of density profiles, the penetration of free polymers into the gel and the connection between the gel and solution phase. The interfacial structure locally equilibrates after roughly 100 chain relaxation times. At late times, the free chains inside the gel undergo a percolation transition if the polymer concentration in the gel exceeds a critical value, which is of the same order as the overlap concentration. The fluctuations of the interface can be described by a capillary wave model that accounts for the elasticity of the gel. Based on this, we extract the interfacial tension of the gel-solution interface. Interestingly, both the interfacial tension and the local interfacial width increase with increasing free polymer concentration - in contrast to liquid-liquid interfaces, where these two quantities are typically anticorrelated.
\end{abstract}

\bigskip

\end{@twocolumnfalse}
]

\section{Introduction}
Hydrogels are three-dimensional macromolecular networks which are able to hold a large amount of water\cite{thakur2018hydrogels}. In recent years, stimuli-responsive hydrogels (smart gels) are attracting increasing interest, especially due to their large potential in  biomedical applications\cite{chen_applications_2019,rajasekar_retracted_2022,WOS:000274422500005, WOS:000957884900001, WOS:000879070100003, WOS:000998044300001}. Such hydrogels respond to external stimuli (changes in pH, ionic strength, temperature, solvent composition) by changing their properties like swelling ratio and elastic modulus. One particularly attractive feature of smart polymers  is reversibility: Induced changes can very often be reversed by simply removing the environmental trigger that caused the response\cite{polybasics,thakur2018hydrogels}. The high sensitivity of hydrogels to small stimuli makes them appealing for a variety of applications, e.g.,  enhanced oil recovery and bio-processing industries, biomimetic actuators, chemical valves and thermoresponsive surfaces \cite{PEPPAS200027, PMID:10407406, QIU2001321}.

The responsive properties of hydrogels in the bulk and at surfaces are typically coupled to each other. For example, one of the most commonly studied thermo-sensitive polymer is  poly(\textit{N}- isopropylacrylamide) (pNIPAAm). pNIPAAm hydrogels in water undergo a volume phase transition from a swollen to a collapsed state if the temperature is increased above 34.3 $^{\circ}$ C \cite{ULLAH2015414,WOS:000329137400010,WOS:000312570500011}. Such swelling-deswelling transitions are accompanied by a change in both gel elasticity \cite{WOS:000323875100018,doi:10.1021/ma00205a036, WOS:000236455200023} and microgel interaction potential \cite{PhysRevLett}. Although the simultaneous change of these properties can be useful for some applications \cite{abc,cell}, it would often be desirable -- e.g., in applications such as switchable cell substrates -- if one could manipulate the gel's elasticity  without affecting its hydrophilicity \cite{micro}. As a strategy to overcome this problem, one of us has recently proposed to use core-shell microgels with a thermo-responsive core and nonthermosensitive shell \cite{core-shellpaper}. We showed that it is possible, using droplet-based microfluidics, to fabricate core-shell microgels with sizes in the range of 60-100 $\mu$m that have a pNIPAAm core and a (thermo-insensitive) polyacrylamide (pAAm) shell. This was achieved by wrapping a pre-synthesized pNIPAAm core by a droplet of semidilute photo-crosslinkable pAAm solution, and later crosslinking the droplet by UV exposure. The properties of the resulting particles depend crucially on the connection between the core and the gel, which is in turn determined by the structure of the interface between the polymer solution and the gel at the time of crosslinking.

From the theory point of view, interfaces between polymer gels and solutions are unusual and fascinating because they combine aspects of liquid-liquid interfaces and elasticity. Computer simulations can give molecular insights into the structure of such interfaces and the mechanisms of solvent diffusion into the gel. While there have been numerous theoretical \cite{D1SM01188J,D1SM00611H,SonuBiswas,NOSELLI201633,LUCANTONIO2013205} and simulation studies on polymer gels\cite{sommer1994structural, everaers1996topological, everaers1999entanglement, rottach2007molecular, mann2005swelling, beyer2022simulations, yang2006montecarlo,brugnoni2018swelling, sliozberg2013effect,fu2014effect,sean2017langevin,anakhov2020stimuli, mourran2016when, tsalikis2023model}, as well as computer models simulating various reaction methods to create micro-gels\cite{YONG2015217,nano11102764,GAVRILOV,articlenumeric, articleMesoscale, doi:10.1021/acs.langmuir.5b03530}, comparatively few studies have been dedicated to gel-liquid interfaces \cite{esteves2013surface, camerin2019migrogels, camerin2020microgels, gumerov2019amphiphilic}. At the same time we are not aware of any work on interfaces between polymer networks and polymer solutions.

In the present paper, we set out to close this gap. We use coarse-grained molecular dynamics simulations based on the Kremer-Grest model \cite{kremergrest, everaers2020kremergres} to study interfaces between a regular polymer network with diamond network topology\cite{ESCOBEDO199985} and semidilute polymer solutions with varying polymer concentration. This choice of network topology was motivated by our experimental work mentioned above \cite{core-shellpaper}, where the core gel was synthesized by free radical polymerization\cite{DASSnew} involving tetrafunctional crosslinkers (\textit{N,N'}-methylenebisacrylamide).
Our model network could also represent tetra-PEG gels\cite{shibayama_exploration_2017,WOS:000381797300188},
which consist of symmetrical tetrahedron-like (four-arm) PEG networks with
excellent physical properties that are attributed to a high homogeneity of the network. \cite{sugimura_mechanical_2013,WOS:000874871800003,WOS:000685071600052,WOS:000294316200028,wang_swelling_2017,WOS:000298198600022,WOS:000362921500062,WOS:000316847500037,WOS:000425473600060}.
To study possible effects of elastic strain, we compare two cases: Slabs of isotropic gels and slabs of anisotropic gels which have been deformed with a ratio very close to 2:1 in lateral direction.

The paper is organized as follows: In section \ref{sec:model} we introduce our simulation model, specify the interaction potentials, and describe the setup and the preparation of initial configurations. The results are presented in section \ref{sec:results}. We first characterize the time evolution of the system after the gel and the solution has been brought into contact, discuss the percolation transition of free chains inside the gel, and then analyze the final equilibrated interface including the capillary wave fluctuations. We summarize and conclude in Section \ref{sec:conclusion}.

\section{Model and Simulation Details}
\label{sec:model}

\begin{figure}[t]
    \includegraphics[width=0.75\columnwidth]
    {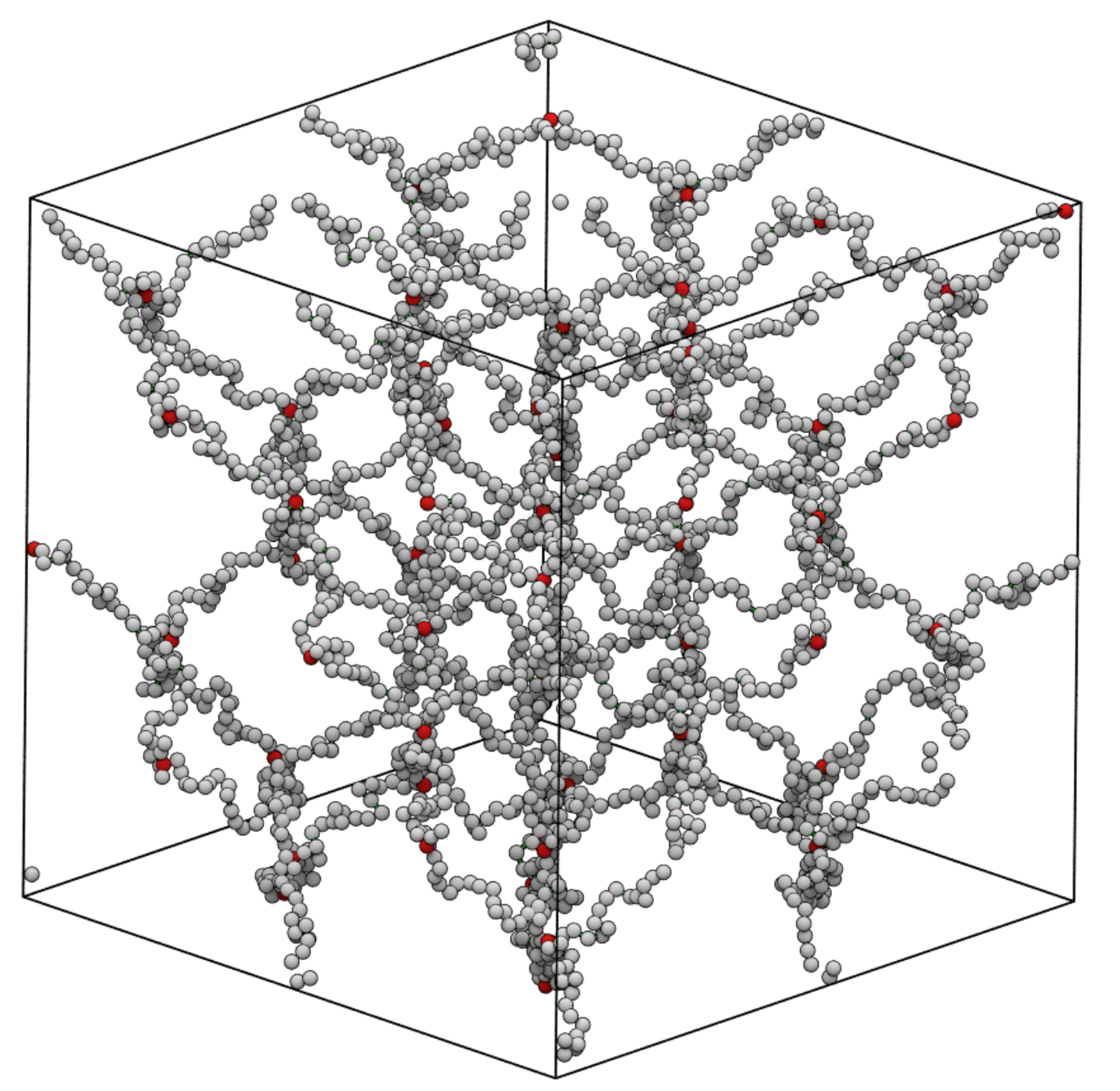}
    \caption{Example of a regular bulk diamond network structure, constructed as described in the main text and not yet relaxed by a NPT simulation. The red beads indicate the cross-linking units, regular beads are grey. The network shown here is crosslinked across periodic boundary conditions in all directions and is smaller than that used in our simulation.
    \label{fig:network}}
    \end{figure}

We consider solutions of $n_l$ linear polymers of length $N_l=102$ beads, which are brought in contact with a polymer gel, using a simple spring-bead model for polymers in implicit good solvent\cite{kremergrest}. The gel is modeled as a regular polymer network with diamond network topology (see illustration in Figure \ref{fig:network})) and consists of $n_s=4464$ strands of length $N_s=102$ beads that are connected by $n_c=2304$ crosslinking beads in total. For simplicity, all beads (''monomers'') are taken to be the same. They interact with purely repulsive Weeks-Chandler-Anderson (WCA) interactions\cite{wcacite} (Equation \protect\eqref{eq:WCA}). In addition, a Finite Extensible Nonlinear Elastic (FENE) potential connects beads that are linked to each other\cite{Jin_2007} (Equation \protect\eqref{eq:FENE}). The corresponding potentials are defined as :
\begin{equation}\label{eq:WCA}
V_{\text{WCA}}(r)= \left\{ \begin{array}{ll}
 4\epsilon \Big[\big(\frac{\sigma}{r} \big)^{12}\!\!\!\! -
   \alpha
   \big(\frac{\sigma}{r}\big)^6 \!\!\! +\frac{\alpha^2}{4}
 \Big]: & \frac{r}{\sigma}< \big(\frac{2}{\alpha} \big)^{1/6}
 \\
 0 & \text{otherwise}
\end{array}
\right.
\end{equation}
\begin{equation}
V_{\text{FENE}}=
\label{eq:FENE}
-\frac{k \:R_{0}^2}{2} \ln{ \Big( 1-\Big(\frac{r}{R_0} \Big)^2 \Big)}
\end{equation}

In the following, we will give all quantities in units of $\sigma$ (length), $\epsilon$ (energy), and $m$ (monomer mass). The basic unit of time is $\tau = \sqrt{m \sigma^2/\epsilon}$.
The interaction strength for FENE bonds is set to $k=30\epsilon/ \sigma^2$ and the maximum extensibility of the bond to $R_0=1.5\sigma$. The parameter $\alpha$ in the WCA potential is chosen  $\alpha=1$ for beads connected by FENE bonds and $\alpha=2$ for all other pairs of beads.

Initial configurations for the simulations are prepared as follows:
To construct the polymer network, $n_c$ cross-linking beads are first placed in the form of a diamond lattice and connected by polymer strands. An example of a resulting small bulk gel structure is shown in Figure \ref{fig:network}. To set up gel slabs, the network is cross-linked across the periodic boundaries in $y$ and $z$ direction, but not in the $x$ direction (different from Figure \ref{fig:network}). Then, to relax the strands, the system is simulated at constant pressure $P=0.001 \epsilon/\sigma^3$ over a time of $8 \times 10^3 \tau$. After the relaxation step, the linear dimensions of the simulation box are $L_x = (232.0 \pm 0.3)\sigma$, $L_y = L_z = (174.1  \pm 0.2) \sigma $.
Next the simulation box is extended in the $x$ direction on both sides of the gel slab
up to a total length of $L_x = 697\sigma$.
To create anisotropic gels, we then apply an affine deformation involving compression
along the $z$ direction and extension along the $y$ direction over a time period of
$10^3 \tau$, and obtain a resized box with linear dimensions $L_z =
102\sigma$ and $L_y=201\sigma$. This deformation step is omitted when preparing isotropic gel slabs. We note that the cross-sectional area of anisotropic gels is a bit lower than that of isotropic gels
($A=L_y \times L_z \sim 2 \times 10^4 \sigma^2$ as opposed to $A \sim 3 \times 10^4 \sigma^2$), which
has a slight influence on some of the results.

The last preparation step consists of filling the empty regions of the extended box with free polymer solution. To this end, $n_l$ fully stretched linear polymer chains are placed along the $z$ direction such that they occupy maximum space. The number $n_l$ varies depending on the concentration of the polymer solution. Then we insert auxiliary hard walls between the gel and the solution regions and carry out NVT simulations of the solution at temperature $k_B T = \epsilon$ over a time $160 \:\tau$ to equilibrate the free chains. The hard walls interact with polymer beads by WCA interactions \protect\eqref{eq:WCA} with size parameter $\sigma_{\text{wall}}=0.1 \sigma$ and $\alpha = 1$. Once the system is equilibrated, we remove the auxiliary walls, reset the clock, and start the actual simulation where the free chains diffuse into the network.

\begin{figure}[t]
\centering
    \includegraphics[width=3.3in,height=1in]{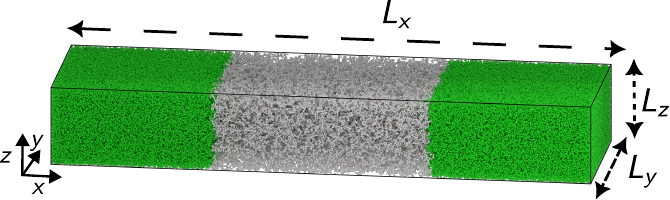}
    \caption{Setup of diffusive interpenetration study: A gel slab (gray) is sandwiched by free polymer chains (yellow) on either sides of it. The free chains permeate into the slab over time. The above snapshot is taken at $800\tau$ at $\rho_{sol_{f}}=0.168 \sigma^{-3}$. The gel is anisotropic.}
    \label{fig2}
\end{figure}

The diffusion simulations (see Figure \ref{fig2}) were done in the NVT ensemble at temperature $k_B T = \epsilon$ using a Langevin thermostat. Typical total simulations times were $1.28 \times 10^6 \tau$, with time step $\Delta t = 0.001 \tau$.  Quantities are averaged over ten independent runs for anisotropic gels and three independent runs for isotropic gels, starting from independently generated initial configurations. All simulations were performed using HOOMD simulation package version $2.9.6$ \cite{Anderson2020} and the simulation snapshots were produced using the fresnel package \cite{fresnel}.

For future reference, we briefly characterize the main properties of the free polymers in our system. The squared radius of gyration of single polymers in dilute solution is $\langle R_g^2 \rangle = (53.3 \pm 0.7) \sigma^2$, and their diffusion constant is $D=(0.010 \pm 0.004 )\sigma^2/\tau$. From these two numbers, one can estimate the characteristic chain relaxation time, i.e., the time it takes the polymer to diffuse over the distance of its own gyration radius, $\tau_d = \langle R_g^2 \rangle/D = 5.3 \times 10^3 \tau$. Below, we will usually present data in terms of rescaled times $t/\tau_d$. From $R_g$, we can also estimate the overlap concentration $\rho^*$, i.e., the concentration where chains start to have significant contact with each other, via \cite{book_doi_edwards} $\rho^* \approx N_l \big/ \frac{4}{3}\pi R_g^3 \sim 0.06 \sigma^{-3}$. Figure\ \ref{fig:Rgvsrho} in SI demonstrates that this value indeed marks the crossover between the dilute and the semidilute regime.
Alternatively, the overlap parameter can also be estimated from the equation of state of the free polymer solution, i.e., the relation between pressure and density.
Fitting this relation to the theoretical expression\cite{book_rubinstein_colby} gives $\rho^* \sim 0.024 \sigma^{-3}$, which is of the same order (see Figure\ \ref{fig:osmoticvsrho} in Supporting Information (SI)).

\section{Results}

\label{sec:results}
The behavior at interfaces is found to be very similar for isotropic and anisotropic gels. Therefore, with few exceptions, we will mostly present the data for anisotropic gels here; additional complementary data for the isotropic gel can be found in the Supporting Information (SI).
We first characterize the evolution of the systems after bringing the gel and the solution into contact, and then analyze the properties of the final interface.

\subsection{Dynamic Evolution of Gel-Solution Interfaces }

\begin{figure}[h!tb]
\centering
    \includegraphics[width=0.8\columnwidth]{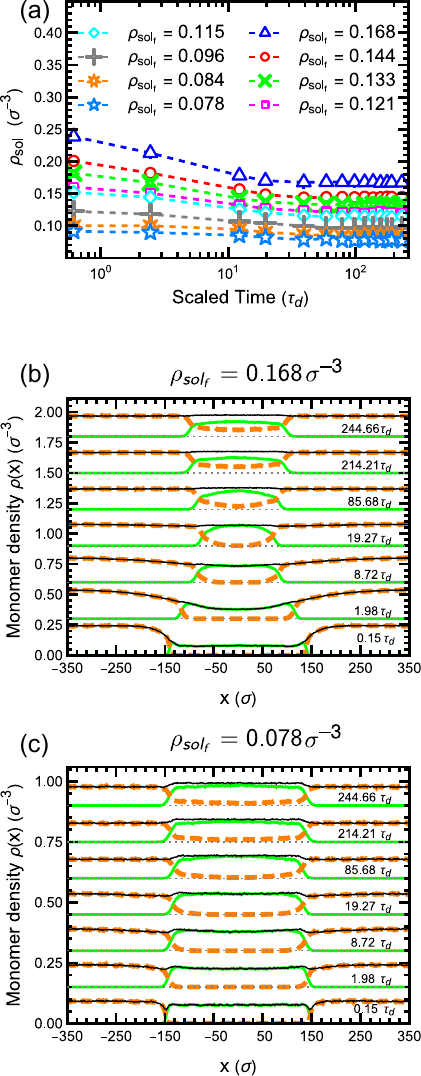}
    \caption{(a) Monomer density in the solution phase versus rescaled time $t/\tau_d$ for different final concentrations $\rho_{\text{sol}_f}$ (in units of $\sigma^{-3}$) as indicated. (b,c): Monomer density profiles across the slab for free polymers (dashed orange), gel strands (solid green), and total (solid black) at different times as indicated for $\rho_{\text{sol}_f} = 0.168 \sigma^{-3}$ (b) and $\rho_{\text{sol}_f} = 0.078 \sigma^{-3}$ (c). Profiles corresponding to subsequent times are shifted upwards by $0.3 \sigma^{-3}$ in (b) and  $0.15 \sigma^{-3}$ in (c). Thin dashed lines show corresponding baselines ($\rho=0$). The gel is anisotropic. See Figure \ref{fig:densprofiso}  in SI for corresponding results for the isotropic gel.}

\label{fig:densprofaniso}

\end{figure}

\subsubsection{Monomer Density Profiles}

After bringing the gel and the polymer solution into contact, polymers diffuse into the gel phase, which somewhat reduces the amount of polymer in the solution phase (see Figure \ref{fig:densprofaniso} (a)). In our simulations, the gel and
the solution initially occupy similar volumes, and reductions of up to 2/3 are observed at the highest concentrations. This could be similar in experimental microfluidics settings if the droplet that serves as precursor of the shell has a similar volume than the pre-synthesized core. In the following, we will
characterize  solutions with different concentrations in terms of the final polymer density in the solution phase $\rho_{\text{sol}_f}$, as calculated at the end of the simulation.

\begin{figure}[t]
\centering
\includegraphics[width=0.8\columnwidth]{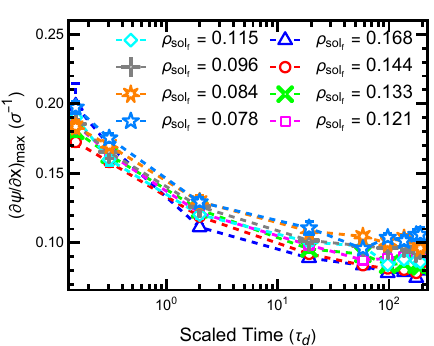}
     \caption{Maximum gradient of the rescaled monomer density difference profiles $\psi(x)$ at the interface (see Equation \protect\eqref{eq:psi}) vs. rescaled time $t/\tau_d$  (orange lines in Fig.\ \ref{fig:densprofaniso}) for solutions with different concentrations $\rho_{\text{sol}_f}$ as indicated. The gel is anisotropic. }
    \label{fig:Sgradienttime:a}
\end{figure}

Figure \ref{fig:densprofaniso} (b),(c) shows monomer density profiles for the gel and the free polymers along the direction perpendicular to the slab (the $x$-direction) for two choices of polymer concentration in the solution. Initially, the polymer solution compresses the gel slab along the $x$ direction.  The compression is caused by the osmotic pressure of the free polymers and is hence more pronounced for higher polymer concentrations. At the same time, polymers start to diffuse from the solution into the gel, and the gel swells.
Both effects counteract each other, and eventually, once sufficiently many  polymers have entered the gel region, the size of the gel increases again. However, the width of the gel slab remains smaller than the original width, due to the fact that the solvent quality of the polymer solution is lower than that of pure (implicit) solvent. The interdiffusion leads to a broadening of the interface between the gel  and the solution. This is demonstrated in Figure \ref{fig:Sgradienttime:a},
which shows the value of the maximum slope of the rescaled difference $\psi$ between the densities of monomers in free polymers ($\rho_{l}(x)$) and in strands ($\rho_{s}(x)$),
\begin{equation}
    \label{eq:psi}
    \psi(x) = \frac{\rho_{l}(x)-\rho_{s}(x)}{\rho_{l}(x)+\rho_{s}(x)}
\end{equation}
It decreases and eventually saturates at times around $t \sim 100 \tau_d$.
We should note that the absolute values of the slopes reported in Figure \ref{fig:Sgradienttime:a} should not be taken literally, since they are affected by capillary wave fluctuations and hence depend on the lateral system size. We will analyze this in more detail further below in Section \ref{sec:width_tension}.

In the concentrated regime, the profile of the final total monomer concentration is roughly constant across the whole system (Figure \ref{fig:densprofaniso} (b), top profile).
At lower concentrations close to the dilute regime, the final total monomer concentration in the gel is slightly higher than in the solution (Figure \ref{fig:densprofaniso} (c), top profile), even if it is initially lower.
This can be explained by the fact that free polymers  gain translational entropy if they enter the gel.

\begin{figure}[h!t]
\centering
\includegraphics[width=0.75\columnwidth]{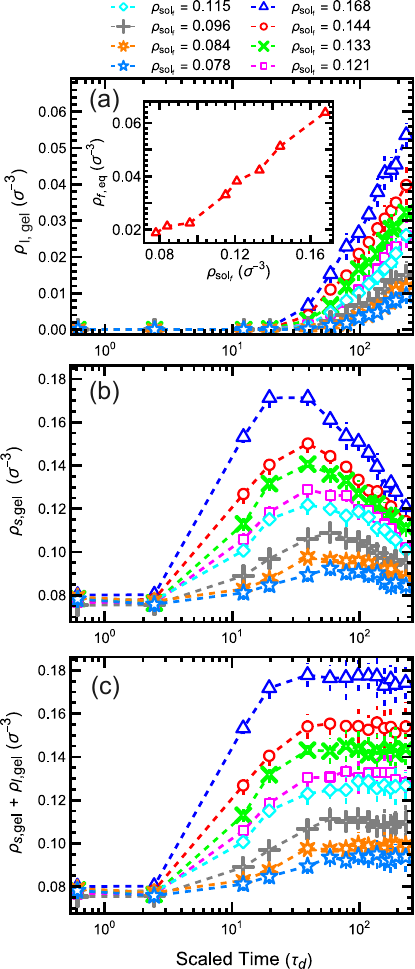}
    \caption{ Monomer densities at the center of the gel slab versus rescaled time $t/\tau_d$ for (a)  monomers belonging to free polymers, (b) gel monomers, (c) all monomers, for different final concentrations in the outer solution  $\rho_{\text{sol}_f}$ (in units of $\sigma^{-3}$) as indicated. Inset in (a) shows equilibrium density of monomers belonging to free polymers inside the slab vs. $\rho_{\text{sol}_f}$, as extracted from a fit of the data in (a) to Eq.\ (\protect\ref{eq:1d_diffusion}).  The gel is anisotropic.}
    \label{fig:figdensmidaniso}
\end{figure}

The monomer densities at the center of the gel are shown as a function of rescaled time for different concentrations $\rho_{\text{sol}_f}$  in  Figure \mbox{\ref{fig:figdensmidaniso}}, separately for
monomers belonging to free polymers (a), monomers belonging to the network (b), and all monomers (c).
The Figure reveals that none of the systems are fully equilibrated as a whole at the end of the simulation:
The monomer density of free chains increases roughly logarithmically at late times, indicating that free polymers continue to diffuse into the gel. This logarithmic increase is compatible with the behavior expected from the solution of the one dimensional diffusion equation inside the gel with Dirichlet boundary conditions (see section \ref{1d_diffusion} in SI),
\begin{equation}
\label{eq:1d_diffusion}
   \frac{\rho_f(0,t)}{\rho_{f,\text{eq}}}  \approx
   \: \frac{1}{2} \Big[
     \tanh \Big( \ln(\frac{Dt}{d^2}) + 2.357 \Big) + 1 \Big],
\end{equation}
where $d$ is the thickness of the gel slab and $D$ the diffusion constant of free polymers. Fitting the data of Figure   \mbox{\ref{fig:figdensmidaniso}}(a) to the approximate expression (\ref{eq:1d_diffusion}), we can estimate the final density $\rho_{f,\text{eq}}$ of monomers from free polymers in the final equilibrated state.
The results are shown in the inset.

The density of gel monomers (Figure \mbox{\ref{fig:figdensmidaniso}}(b)) first increases due to the initial compression the gel, and then decreases again as soon as the free polymers start moving in. Interestingly, the total density at the center of the gel saturates at a roughly constant value at times around $40-80 \tau_d$. The predominant factor setting the overall density within the gel is the osmotic pressure exerted by the surrounding polymer solution on the polymer/gel interface. It depends on the densities of free polymers in the vicinity of the interface. Our findings thus suggest that the interface equilibrates much faster than the entire gel, and that it will be possible to analyze the local properties of interfaces.

\subsubsection{Connection between gel and solution}

\begin{figure}[t]
\centering
    \includegraphics[width=0.75\columnwidth]{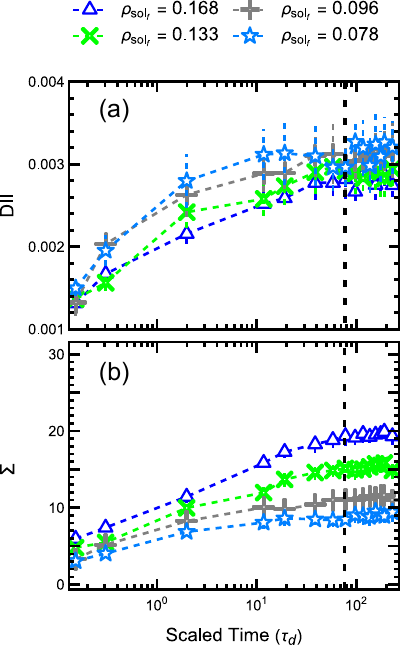}
    \caption{(a) DII vs scaled time $t/\tau_d$ for various concentrations $\rho_{\text{sol}_f}$.  (b) Scaled areal density $\Sigma$ of free chain monomers with connection to the interface vs. scaled time $t/\tau_d$. Dashed line indicates time after which $\Sigma$ saturates. The gel is anisotropic.}
    \label{fig:dii_aniso}
\end{figure}

Next we characterize the interpenetration between free polymers and gel strands,
which determines the strength of the connection between the two phases after
crosslinking. As one measure of interpenetration, we evaluate the ''Degree of
Interfacial Integration'' (DII), which has been introduced by Shi et al \cite{interfdiff}
in studies of multilayer polymer films. It is defined as
\begin{equation}
    \text{DII} = I /I_{\text{max}} \quad \text{with} \quad
    I = \langle d \rangle  \cdot \langle S \rangle
\label{eq:dii}
\end{equation}
and
\begin{equation}
    S = \frac{n_{_{li}}}{A}, \quad
    d = \min\big(\sum_{j: x_j < 0}|x_j|, \sum_{j:x_j > 0} x_j \big).
\end{equation}
Here $A=L_y \cdot L_z$ is the area of the interface and $n_{_{li}}$ is the number of free polymers crossing the interface. The quantity $d$ is evaluated for each of these polymers separately. The sum $j$ runs  over all monomers and $x_j$ is the $x$-position of the monomer with respect to the position of  the interface. Hence $S$ characterizes the areal density of free polymers stitching through the interface, and the ''depth'' of the stitch, $d$, is largest if the polymer is fully stretched perpendicular to the interface with half of its monomers being on one side and the other half on the other side. The quantity DII is defined as the ratio of $I$ and its
maximal possible value $I_{\text{max}}$ for given polymer concentration $\rho_{\text{sol}_f}$ in solution, i.e.,
$I_{\text{max}} = d_{\text{max}} \cdot S_{\text{max}}$ with

\begin{equation}
    S_{\text{max}} = \rho_{\text{sol}_f} \: r_0,  \qquad
    d_{\text{max}} = r_0 \: N_l \: (N_l+2) \: /8,
\label{eq:diimax}
\end{equation}
where $r_0 = 0.97 \sigma$ is the equilibrium bond length. This maximum value is reached in the very hypothetical case that all free chains are fully stretched and oriented perpendicular to the surface. To get an estimate of a realistic range of $I$, we can also estimate the value of $I$ for randomly distributed Gaussian coils with gyration radius $R_g^2 = r_0^2 N_l/6$. In $x$ direction, the monomers of such a coil with center of mass at $x_{\text{cm}}$ are distributed according to $P(x) \sim \exp(-3(x-x_{\text{cm}})^2/2 R_g^2)$. Introducing a hypothetical cutoff $R_c$ for the maximum value of $|x-x_{\text{cm}}|$,  and after some math, one gets
\begin{equation}
    \langle S_{\text{Gauss}} \rangle = \frac{\rho_{\text{sol}_f}}{N_l}\: 2 R_c,  \quad
    \langle d_{\text{Gauss}} \rangle = N_l \: \frac{1}{12} \frac{R_g^2}{R_c} ,
\label{eq:iicoil}
\end{equation}
in the limit of large  $R_c$ ($R_c\gg 10 R_g$). The DII for such a solution of noninteracting Gaussian coils would thus be given by
\begin{equation}
    \text{DII}_{\text{Gauss}} = \frac{I_{\text{coil}}}{I_{\text{max}}}
    = \frac{2}{9} \: \frac{1}{N_l+2}
     \approx 0.0024
\label{eq:diicoil}
\end{equation}
for chains of length $N_l = 102$.

\begin{figure}[t]
\centering
    \includegraphics[width=0.75\columnwidth]{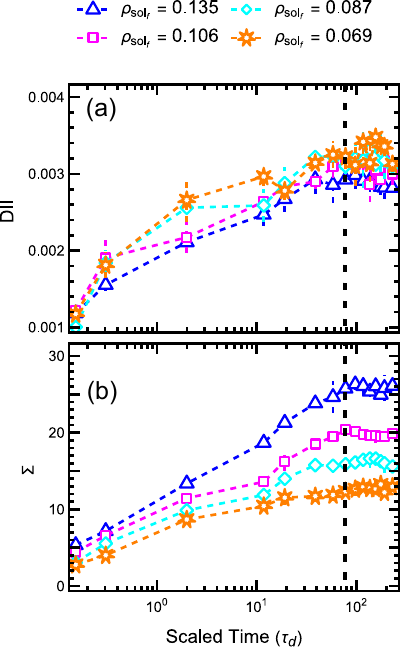}
    \caption{Same as Figure \protect\eqref{fig:dii_aniso} for the isotropic gel}
    \label{fig:dii_iso}
\end{figure}

The evolution of the DII with time in our system is shown in Figure \ref{fig:dii_aniso} (a) for anisotropic gels and \ref{fig:dii_iso} (a) for isotropic gels. The behavior in both cases is very similar. It increases and then saturates at a value around DII=0.003, which is slightly higher than the hypothetical value for randomly distributed Gaussian coils. Initially, the DII rises more rapidly if the solution is less concentrated, but the final value is independent of $\rho_{\text{sol}_f}$ within the error. We should note that the absolute number of ''stitches'' of course increases with  $\rho_{\text{sol}_f}$, therefore the figure also tells us that the strength of the connection between the two phases increases with increasing polymer concentration in the solution.

As a second measure to characterize the interface, we define the quantity $\Sigma$
\begin{equation}
\Sigma = \langle N_{li,\text{gel}} \rangle \cdot \langle S \rangle\: \langle R_g^2 \rangle,
\end{equation}
where $\langle N_{li, \text{gel}}\rangle$ counts the number of monomers of a free polymer inside the gel, averaged over all polymers that cross the interface and $\langle R_g ^2 \rangle$ is the mean squared radius of gyration of a chain in a dilute solution. $\Sigma$ can be seen as number of monomers inside the gel  within a lateral area $R_g^2$ that have some connection to the solution phase. This quantity is shown as a function of time in Figures \ref{fig:dii_aniso} (b) and \ref{fig:dii_iso} (b). It rises initially in a roughly logarithmic fashion (linear in a log-linear representation) and then levels off. Comparing Figure \ref{fig:dii_aniso}  with Figure \ref{fig:Sgradienttime:a}, we notice that both, the DII and $\Sigma$, saturate at about the same time as the local monomer density profiles. The saturation time can be as long as $t_{\text{sat}} \sim 100 \tau_d$. At late times beyond $t_{\text{sat}}$,  the local density profiles and the local connectivity structure at the interface no longer change. We infer that for times $t > t_{\text{sat}}$, the interfaces can be considered to be at local equilibrium, even though polymers keep diffusing into the gel according to Figure \ref{fig:figdensmidaniso}.

We conclude that the best strategy to fabricate core-shell particles with given (pre-determined) connection between core and shell, using the droplet-based method sketched in the introduction, is to adjust the concentration of polymers in the solution. Unfortunately, this will likely also affect the structure of the shell network after irradiation, e.g., the density of crosslinking  points. An alternative strategy is to adjust the time of crosslinking. Weak connectivities can be achieved by crosslinking shortly after the first polymer-core contact, and stronger connectivities will result if the crosslinking time is chosen in the long-time limit.

\subsubsection{Percolation  of free chains inside the gel}

\begin{figure*}[tbh]
    \includegraphics[scale=0.9]{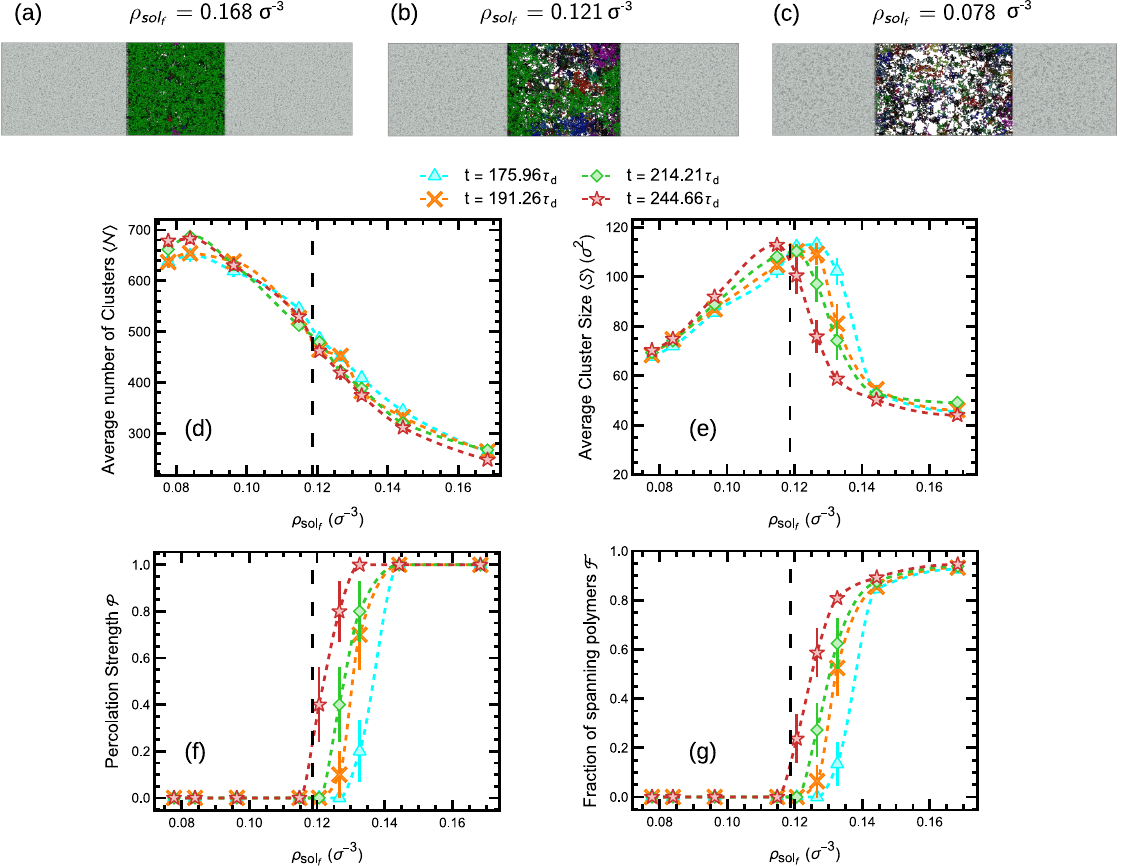}
    \caption{(a-c) Cluster formation of free chains within the gel region as viewed from top of the box (along z-axis) at different final polymer densities $(\rho_{sol_f})$ as indicated. Different clusters are colored differently in order to be distinguishable. The light gray color indicates solution region. (d-e) Characteristics of cluster distributions vs.\ $(\rho_{sol_f})$ at different times as indicated: (d) Average number of clusters; (e) Average cluster size (mean squared radius of gyration ) of finite clusters; (f) Percolation strength (see text); (g) Fraction of free polymers inside gel that are part of the spanning cluster. The gel is anisotropic.}
   \label{fig:percAnsio}
\end{figure*}

If the concentration of polymers in solution is sufficiently high, one notices that the free polymers in the gel start to come in contact with each other and form clusters. Figure \ref{fig:percAnsio} (a)-(c) shows examples of final configuration snapshots for different concentrations $\rho_{\text{sol}_f}$. Only the free polymer chains are shown, the gel region is singled out, and inside this region,  chains belonging to different clusters are colored differently. Here, chains are taken to belong to the same clusters if at least two monomers from different chains have a distance less than $2 \sigma$. The cluster analysis was performed using the freud-analysis package \cite{freud2020}. At low concentration ($\rho_{\text{sol}_f} = 0.078 \sigma^{-3}$), some clusters form, but they are isolated. At higher concentration ($\rho_{\text{sol}_f} = 0.121 \sigma^{-3}$), a spanning cluster emerges that connects both sides of the slab. At the highest concentration ($\rho_{\text{sol}_f} = 0.168 \sigma^{-3}$), the spanning cluster contains almost all free chains in the gel.

These findings suggest the existence of a percolation transition  at some threshold density $\rho_{sol_{f}}^c$. The percolation transition is a well-known geometric transition \cite{WOS:000073152400005,2010M2009329,WOS:000294196500008,WOS:000682785900121, DEFREITAS199981,WOS:000357230800012,WOS:000234617600059} associated with critical exponents and universal power laws, which can be extracted, e.g., by performing a finite size analysis. Unfortunately, carrying out such an analysis is not possible for our system, because even the final configurations are not yet fully equilibrated as discussed earlier. Nevertheless, we can discuss typical aspects of percolation and monitor the onset of percolation with time.

Figure  \ref{fig:percAnsio} (d)-(e) shows a selection of quantities that characterize the distributions of clusters of free chains in the gel for different times, and  as a function of the polymer concentration in solution, $\rho_{\text{sol}_f}$. The first quantity, shown in Figure \ref{fig:percAnsio} (d), is the average number of independent clusters  $\langle \mathcal{N} \rangle$. The graph confirms the discussion above. At lower concentrations, the gel contains many scattered small clusters, which are not connected to each other. With increasing concentration, clusters start to merge and the total number of clusters decreases. The quantity $\langle \mathcal{N} \rangle$ decreases smoothly as a function of $\rho_{\text{sol}_f}$ and shows no signature of a  transition.

Next we plot the average cluster size $\langle \mathcal{S} \rangle$ of finite clusters as a function of $\rho_{\text{sol}_f}$ in Figure \ref{fig:percAnsio} (e). Finite clusters are the ones which are not part of a spanning cluster. At lower concentrations, several isolated finite clusters coexist, and therefore, the average size ($\langle \mathcal{S} \rangle$) of the clusters is small. With increasing $\rho_{\text{sol}_f}$, clusters merge and grow in size, as can also be seen in the snapshots, Figure \ref{fig:percAnsio} (a), (b). Above the percolation threshold, more and more larger clusters merge with the spanning cluster. The remaining clusters are small, hence $\langle \mathcal{S} \rangle$ decreases beyond the threshold concentration.

Figure \ref{fig:percAnsio} (f) shows the so-called percolation strength $\mathcal{P} = \langle \chi \rangle$ , where $\chi=1$ if a configuration contains a spanning cluster, and $\chi = 0$ otherwise. The curves for  $\mathcal{P}$ versus $\rho_{\text{sol}_f}$ strongly suggest the existence of a percolation transition, i.e., a jump from $\mathcal{P}=0$ to $\mathcal{P}=1$ at $\rho^c_{\text{sol}_{f}} \sim 0.12 \sigma^{-3}$. The jump is smoothed out due to the finite size of the system. The   onset of percolation, i.e., the concentration $\rho_{\text{sol}_f}$ where percolation is observed, decreases with increasing time. This can be explained if we assume that percolation is only possible once the concentration of free chains inside the gel exceeds a certain value. Comparing Figure \ref{fig:percAnsio} (e) with Figure  \ref{fig:figdensmidaniso}, we can estimate the threshold concentration inside the gel, $\rho^c_{l,\text{gel}}$,
where percolation sets in. The data suggest $\rho^c_{l,\text{gel}} \sim 0.03 \sigma^{-3}$, which is of the same order as the overlap concentration $\rho^*$. This seems low, but not unreasonable. Experiments have shown that the sol-gel transition in solutions of crosslinkable polymers -- a transition similar to percolation -- may set in at polymer concentrations as low $\rho^*/6$\cite{Sakai2016}.

Finally, we consider the fraction of free polymers in the gel that belong to the spanning cluster,
$\mathcal{F}$. This quantity would typically be used as an order parameter for the transition. It is shown as a function of $\rho_{\text{sol}_f}$ in Figure \ref{fig:percAnsio} (g). The curves show that $\mathcal{F}$ increases beyond the transition, but less sharply than $\mathcal{P}$, indicating that the transition might be continuous in the limit of infinite systems.

In this section, we only show the data for anisotropic gels; the corresponding data for isotropic gels can be found in SI, section \ref{SI:isoPerc}. They are qualitatively similar, but percolation sets in at lower concentrations $\rho_{\text{sol}_f}$ than in the anisotropic gel. We attribute this to the fact that our isotropic gels have a higher cross section area $A$ and are hence
less dense than the anisotropic gels, therefore they can be more easily penetrated by free polymers. The density of monomers belonging to free polymers inside the gel at the percolation threshold is around $\rho^c_{l,\text{gel}} \sim 0.03 \sigma^{-3}$, as in the anisotropic gel.

From a practical point of view, our results on the percolation transition of shell polymers inside the core might be relevant because they give insight into the connection between the core and shell beyond the direct interfacial region. Submillimeter size core gels will typically not be fully invaded by shell polymers during the preparation time. However, once the shell polymer concentration exceeds $\rho^*/2$ in a region close to the core surface, shell polymers may form large interconnected clusters which can be crosslinked and will further strengthen the connection between core and gel. The thickness of the region where these clusters occur depends upon the time duration during which shell chains are permitted to interdiffuse prior to being crosslinked.
\subsection{Structure of equilibrated interfaces}

Having studied the time evolution of the system after bringing the core gel and shell solutions  into contact, we will now investigate the properties of the final, locally equilibrated interfaces in more detail.

\subsubsection{Local monomer motion and chain conformations}

\begin{figure}[!t]
\centering
    \includegraphics[width=\columnwidth]{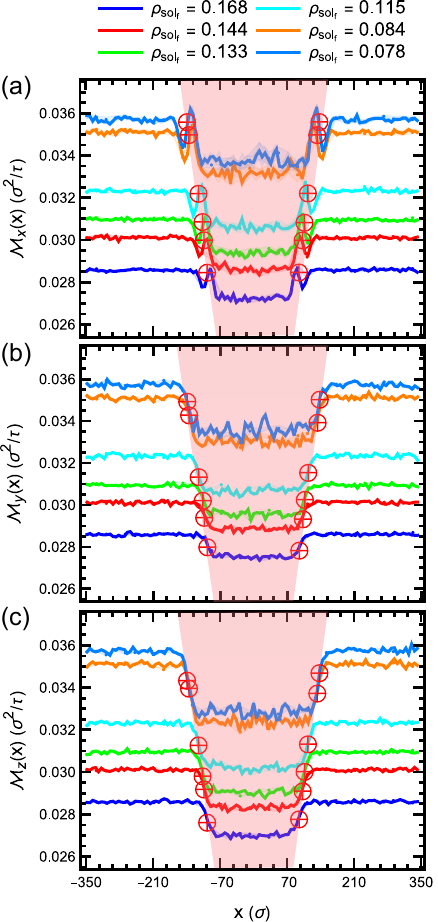}
\caption{ Profiles of monomer motion  $\mathcal{M}_{_\alpha}(x)$  in $\alpha=x$ (a), $\alpha=y$ (b) and $\alpha=z$ (c) direction vs. $x$-coordinate of monomers. The red symbols $\bigoplus$ indicate the approximate position of interface, the red shading the gel regions. The gel is anisotropic.}
\label{fig:anisodiff}
\end{figure}

To characterize the local motion of monomers, we calculate their mean squared displacements from every consecutive frames and define the quantity:
\begin{equation}
         \mathcal{M}_{_\alpha}(x)=\frac{ \langle (r_{_\alpha}(t+\Delta t_s) - r_{_\alpha}(t))^2\rangle}{\Delta t_s}.
     \label{eq: mobilitymeasureeq}
 \end{equation}
Here $\alpha=x,y,z$, $r_{_\alpha} (t)$ are the $\alpha$th component of the position of the monomers at time $t$ , and $\Delta t_s=800 \tau$ refers to the time interval between two consecutive frames /snapshots.

 Figure \ref{fig:anisodiff} shows the profiles of $\mathcal{M}_i$ across the slab  for anisotropic gels. The corresponding results for isotropic gels can be found in SI, Figure \ref{fig:isodiff}. As one might expect, the local motion is reduced inside the gel, compared to the solution phase, due to the friction with the network. Furthermore, we also see that $\mathcal{M}_{_\alpha}$ decreases with increasing $\rho_{sol_f}$ due to the friction with other free polymers. Inside the gel, monomers move more in $y$ direction than in $z$ direction, which is a consequence of the anisotropy of the gel. The difference disappears in isotropic gels as shown in SI, Figure \ref{fig:isodiff}. In the $x$-direction, the value of $\mathcal{M}$ inside the gel is intermediate between the values in $y$ and $z$ directions (identical in isotropic gels). More interestingly, the profile of $\mathcal{M}_x(x)$ features an oscillation close to the interface. Monomers just outside the gel move less than their neighbor monomers, since the gel surface acts as a barrier to particle motion. Monomers just inside the gel move more than their neighbors, since their moves take them into the gel-free region. The oscillation helps to maintain net zero monomer flux across the interface: Monomers just outside of the interface cross the interface less often, however, they have higher density. Monomers just inside cross the interface more often, but their density is lower.

\begin{figure}[!t]
\centering
    \includegraphics[width=1.05\columnwidth]{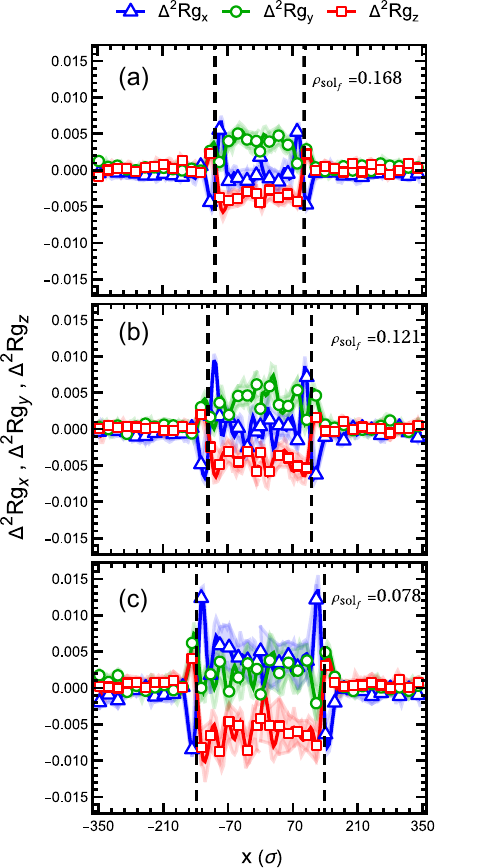}
  \caption{Profiles of chain conformation parameters $\Delta^2 {R_g}_{_\alpha}$ (see text) across the gel slab for different concentrations  $\rho_{\text{sol}_{f}}$ as indicated. The $x$ coordinate refers to the $x$ center of mass of chains. Black dashed line indicates the position of the interface, shaded bands represents the error.  The gel is  anisotropic}
\label{fig:anisoconf}
\end{figure}

Next we examine the conformations of the free polymers as a function of the x-position of the simulation box. Following Adhikari et al \cite{naradhikariRg, adhikarianother}, we define anisotropy parameters of the gyration tensor as:

\begin{equation}
         \Delta^2 {R_g}_{_\alpha} = \frac{3 \langle {R_g}_{_\alpha} ^2 \rangle - \langle {R_g}^2 \rangle}
           {2 \langle R_g^2 \rangle},
     \label{fig:rgconfeq}
 \end{equation}
 where $\alpha=x,y,z$ as before. Here $\langle R_g^2 \rangle \sim 42-47 \sigma^2$
 is the mean squared radius of gyration of free chains and $\langle {R_g}_{_\alpha} ^2\rangle$ the corresponding component in the direction $\alpha$.
 The value of $ \Delta^2 {R_g}_{_\alpha}$ is positive if chains orient or are elongated along the $\alpha$ direction, and negative if the chain orientation is orthogonal to that direction or if chains are squeezed. If  $ \Delta^2 {R_g}_{_\alpha}$ is zero for all $\alpha$, the distribution of chain orientations and conformations is isotropic.

The resulting profiles for three different concentration values $\rho_{sol_{f}}$ in solution
are shown for anisotropic gels in Figure \ref{fig:anisoconf}. Here the $x$ coordinate of a chain is taken to be the position of its center of mass.  In the solution, the values of $ \Delta^2 {R_g}_{_\alpha}$ are zero for all $\alpha$, indicating that the chain conformations are isotropic. Inside the gel, the conformations become anisotropic: $ \Delta^2 {R_g}_{_y}$ is positive and $ \Delta^2 {R_g}_{_z}$ is negative, indicating that the chains align in the direction in which the anisotropic gel has been stretched. Indeed, in the isotropic gel, these effect disappears and both $ \Delta^2 {R_g}_{_y}$ and $ \Delta^2 {R_g}_{_z}$ are zero within the error (see Figure \ref{fig:isoconf} in SI). The value of $\Delta^2 {R_g}_{_x}$ inside the gel is mostly zero, except at the lowest concentration, where it is positive, most likely due to the swelling of the gel.

Close to the interface, the profile of the perpendicular component of the gyration tensor, $\Delta^2 {R_g}_{_x}$, exhibits an oscillation with a negative dip on the solution side and a positive dip on the gel side. This indicates that chains in solution are slightly squeezed close to the gel interface, similar to chains close to surfaces. On the gel side, the chains are slightly elongated due to a tendency to extend loops or chain ends into the solution region.

\subsubsection{Interfacial width and interfacial tension}

\label{sec:width_tension}

Finally, we discuss the fluctuations of the gel/solution interface. At finite temperatures, interfaces between liquids are not perfectly flat. They undulate due to thermal fluctuations. These undulations, also called capillary waves, significantly broaden the apparent interfacial profiles, such that the apparent width $w$ grows indefinitely with the lateral system size $L_\parallel$, following $w^2 \sim L$ in two dimensions and $w^2 \sim \ln(L)$ in three dimensions \cite{2dinterf,3dinterf,book_widom_rowlinson}. For interfaces between demixed phases in homopolymer blends, the phenomenon  has been studied intensely  \cite{werner1999intrinsic,width2,width3,carelli2005approaching}, and it has been shown that a detailed analysis of the interfacial width broadening can be used to extract the interfacial tension. Our goal here is to use a similar approach to calculate the interfacial tension of gel/solution interfaces.

However, the situation is complicated by the elasticity of the gel. First, the notion of an interfacial tension is somewhat ambiguous when dealing with elastic interfaces. It can be defined in two different ways, i.e., (i) as the excess free energy per area at the surface, or (ii) as the work per area required to stretch the surface by a given amount. The first definition, also referred to as ''surface energy'' ($\sigma$), describes the free energy increase if the surface is enlarged by creating more interface, e.g., by cutting the gel up. The second definition, also referred to as ''surface stress / tension'' ($\gamma$), is related to the mechanical force that opposes stretching of the gel. At liquid/liquid interfaces, the two quantities are the same. At interfaces involving elastic materials, they are different, and related to each other via the so-called Shuttleworth relation \cite{shuttleworth1950thesurface, weijs2014capillarity, andreotti2020statics}
\begin{equation}
\gamma = \sigma + A \: \frac{\partial \sigma}{\partial A}.
\end{equation}
In the context of capillary wave fluctuations, interface fluctuate without adding/removing crosslinks or strands in lateral direction, hence the relevant quantity is the surface stress $\gamma$.

A second complication arises from the fact that interface fluctuations are not only penalized by interfacial stress, but also by the elastic distortion of the gel perpendicular to the interface. Therefore, the standard capillary wave formalism must be modified accordingly.

Capillary wave fluctuations are typically described in terms of effective interface Hamiltonians. In our system, we have two interfaces, corresponding to the two surfaces of the slab. We assume that they are sufficiently far apart such that the
coupling across the gel can be neglected. Hence every surface can be described by an independent two-dimensional manifold which is coupled to an elastic medium. Neglecting overhangs, we describe the surface by a function $h(y,z)$ corresponding to its local position in $x$ direction relative to its mean position. For simplicity, we ignore the fact that the interfacial tension $\gamma$ might be anisotropic for anisotropic gels.

The linearized interface Hamiltonian then reads

\begin{multline}
   \label{eq:h}
    \mathcal{H}[h] = \int \!\! \text{d}y \: \text{d}z \Biggl[\gamma \left(1+\frac{1}{2}\biggl(\frac{ \partial h}{ \partial y}\right)^2\\ + \: \frac{1}{2}\left(\frac{ \partial h}{ \partial z}\right)^2\biggr)  +\frac{1}{2}B \: h^2\Biggr]
\end{multline}
Here $B$ is an elastic coupling constant.
Applying a two dimensional Fourier series expansion
\begin{equation}
\label{eq:ft}
    h(z,y)  = \frac{1}{\sqrt{A}} \:
  \sum_{\vec{q}}\widetilde{h}(\vec{q})\exp(-i {\vec{q}}\cdot \vec{r})
\end{equation}
to Equation \protect\eqref{eq:h} and then exploiting the generalized equipartition theorem, we can calculate the thermally averaged height correlations
of the interface $h$ in Fourier space as:
\begin{equation}
\label{eq:hfluc}
\langle |\widetilde{h}(\vec{q})|^2 \rangle
   =\frac{k_BT }{(B+\gamma q^2) },
\end{equation}
Now we use equation \protect\eqref{eq:hfluc} to find the variance $\langle h^2 \rangle$ of the distribution of interface positions ${\cal P}(h)$ in a patch with lateral size $L_{\parallel}$. We assume that the apparent monomer density profile $\rho(z)$ at the interface can be written as the convolution of an ''intrinsic'' profile with ${\cal P}(h)$
\begin{equation}
\rho(z) = \int \: \text{d}h \: {\cal P}(h) \: \rho_{\text{int}}(z-h)
\end{equation}
The interfacial broadening can then be approximated by  \cite{semenov1993theory,werner1997anomalous}
\begin{equation}
    w^2 = w_{\text{int}}^ 2 +  \frac{\pi}{2} \langle h^2 \rangle,
\end{equation}
 resulting in
\begin{equation}
\label{eq:width}
w^2=w_{\text{int}}^2
 +\frac{k_BT}{8 \gamma}\ln{\left(\frac{1+g b_{\text{int}}^2}{1+gL^2_{||}}\right)}
 +\frac{k_BT}{4\gamma}\ln{\left(\frac{L_{\parallel}}{b_{\text{int}}}\right)}
\end{equation}
where $g={B}/{4 \pi^2\gamma}$
and $b_{\text{int}}$ is an intrinsic length scale. Details of the derivation can be found in the Appendix.

\begin{figure}[t]
\centering
    \includegraphics[width=.9\columnwidth]{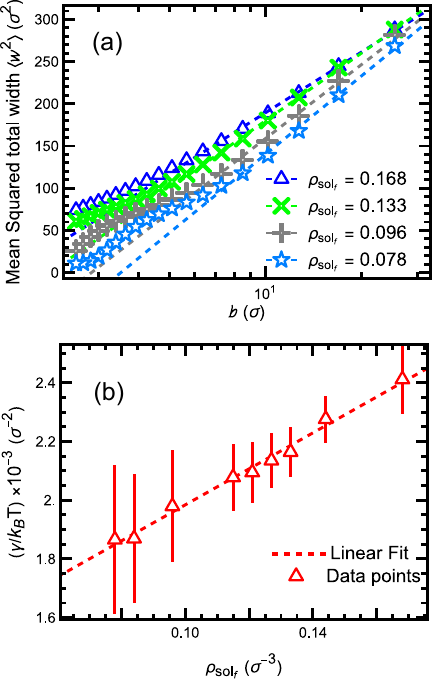}
\caption{(a) Apparent squared interfacial width $w^2$ as a function of block size $b$ (see text for explanation).  The dashed lines represent the fit to Equation \protect\eqref{eq:width}, see text for details. (b) Fitted value of interfacial tension $\gamma$ vs polymer concentration $\rho_{\text{sol}_f}$.
 The dashed line shows a linear fit to the data. The gel is anisotropic.
}
\label{fig:anisowidthtension}
\end{figure}

\begin{figure}[t]
\centering
    \includegraphics[width=0.9\columnwidth]{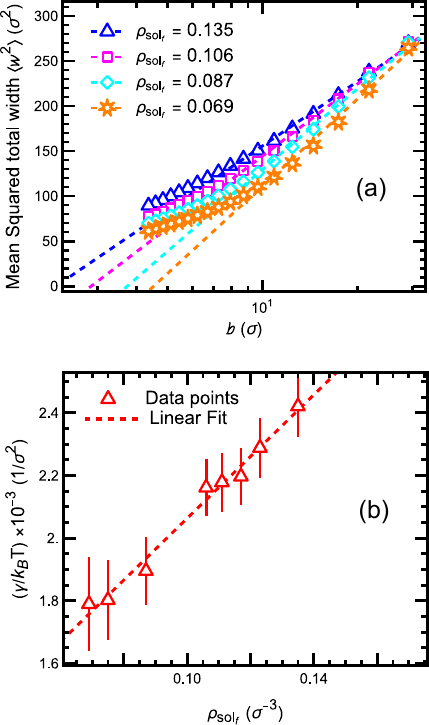}
    \caption{Same as Figure \protect\ref{fig:anisowidthtension} for the isotropic gel.}
\label{fig:isowidthtension}
\end{figure}

We now apply the relation \protect\eqref{eq:width} to analyze the gel-polymer interfaces. To this end, we divide the $(L_y, L_z)$ surface of the simulation box into equal number of bins \cite{werner1999intrinsic,actpasInt,mueller_schmid_review}, such that we obtain $n^2$ sub-blocks with dimensions $b\times b \times L_x$. We determine the density profile within each of these sub-blocks, determine the interfacial width in the sub-block as described in Appendix \ref{appendix:graph}, and average over all sub-blocks and all configurations. The results are shown as a function of sub-block size in Figures \ref{fig:anisowidthtension} (a) and \ref{fig:isowidthtension} (a) for the anisotropic and isotropic gel, respectively. For large sub-block sizes, the data are well described by Equation \protect\eqref{eq:width} (see below for details on the fitting procedure). At lower block sizes around $b \leq 10 \sigma$, the curves deviate from the theory and bend upwards. In this regime, $b$ becomes comparable to the radius of gyration of strands, and hence the description of the interface in terms of a simple interface Hamiltonian is bound to fail. The curves for the anisotropic gel exhibit some substructure for lower $\rho_{\text{sol}_f}$, see Figure \ref{fig:anisowidthtension} (a). This substructure is not observed in isotropic gels Figure \ref{fig:isowidthtension} (a), indicating that the local interface structure is affected by gel anisotropy.

We note that it is not possible, just from looking at Figures \ref{fig:anisowidthtension} (a) and \ref{fig:isowidthtension} (a), to extract an obvious ''intrinsic'' block size $b_{\text{int}}$ which could be used to unambiguously determine the intrinsic width $w_{\text{int}}$. A similar observation was made in interfaces of homopolymer blends \cite{werner1999intrinsic}.

We have used Equation \protect\eqref{eq:width} to fit the interfacial width data for large block sizes $b$. The least square fitting procedure resulted in two different types of fits: one giving large values of the elastic parameter $B$, $B > 0.5 \: \epsilon/\sigma^4$ or higher, and one giving values of order $B \sim  10^{-9} \: \epsilon/\sigma^4$, i.e., $B$ is zero within the error. Considering that the bulk modulus of bare gels is small (see SI, Figure \ref{fig:modulus}, we consider the results of the first type of fit to be physically unreasonable, and selected the second type. The fitted values for $w_{\text{int}}^2$ and $b_{\text{int}}$ had large errors, because they strongly depend on each other. However, if we applied constraints on one of these parameters, it only affected the other, and not the fit values of $B$ and $\gamma$. Therefore, we infer that the fit values for  $\gamma$ are reliable, despite the fact that applying a four-parameter-fit to the curves in Figures \ref{fig:anisowidthtension} (a) and \ref{fig:isowidthtension} (a) seems somewhat brash.

The results for the interfacial tension, $\gamma/k_BT$, are shown as a function of the concentration of polymers in solution, $\rho_{\text{sol}_f}$, in Figures \ref{fig:anisowidthtension} (b), \ref{fig:isowidthtension} (b), respectively. The  interfacial tension increases roughly linearly with increasing concentration $\rho_{\text{sol}_f}$. It grows a bit faster for the isotropic gel, which we attribute again to the fact that our isotropic gels are less dense, hence the interfaces  at the same bulk concentration $\rho_{\text{sol}_f}$ contain a higher amount of free polymers. More importantly, Figures \ref{fig:anisowidthtension} (a) and \ref{fig:isowidthtension} (a) clearly show that the ''intrinsic width''  itself, no matter how we define it -- i.e. for which intrinsic block size $b_{\text{int}}$ we evaluate it -- also increases with increasing concentration $\rho_{\text{sol}_f}$.

The positive correlation between the interfacial width and the interfacial tension stands in stark contrast with the conventional behavior seen at liquid-liquid interface, where these two quantities are typically anticorrelated: The interfacial tension increases with increasing incompatibility of the two components, whereas the interfacial width decreases. In Cahn Hilliard-type theories, the interfacial tension and the interfacial width are in fact predicted to be inversely proportional to each other. However, this does not hold true at the gel/solution interface, underscoring its peculiar nature. In our model system, the gel strands and the polymers in solution are not incompatible but identical. The origin of  interfacial tension lies in elasticity and entropy. As the polymer concentration  $\rho_{\text{sol}_f}$ increases, the polymers swell the gel, leading to an increase of the interfacial width. On the other hand, the swollen gel stiffens, which also increases the interfacial tension.

\section{Conclusions and Outlook}

\label{sec:conclusion}

In the current study, we have investigated the interdiffusion of polymers from solution into regular polymer networks. Even though the study was motivated by the practical problem of understanding physical processes during the preparation of core-shell particles by microfluidics, it also offers general insights into the structural properties of interfaces between polymer gels and solutions and the dynamics of interdiffusion.

Comparing the gel/solution interfaces to liquid-liquid interfaces between demixed homopolymer phases, we find a number of qualitative differences. First, the evolution of the interfacial region at the onset of interdiffusion is nonmonotonic due to interplay of two competing effects: The compression of the gel due to the osmotic pressure of the polymers in solution, and the swelling of the gel due to the penetration by polymers. As a result, the gel first shrinks and then expands again. A second remarkable difference is a positive correlation between the interfacial tension (the surface stress) and the interfacial width, which stands in contrast to interfaces between immiscible liquids, where these two quantities are inversely correlated.

From a practical point of view, our study can offer insights that might help to optimize experimental strategies for preparing multiphase polymer hydrogels by sequential crosslinking of layers. In good solvent, the connectivity between layers, as characterized by the ''degree of interfacial integration'' \cite{interfdiff}, saturates at values that are characteristic for randomly distributed Gaussian chains. However, we find that it takes a long time until saturation is reached. Even though entanglement effects are presumably not important in our system -- even in  dense melts, the entanglement length of flexible Kremer-Grest chains is of order $N_e \sim 50 $ beads \cite{svaneborg2020characteristic} -- the saturation time is found to be orders of magnitude higher than the chain relaxation time. This suggests that it should be possible to tune the connectivity between the layers by tuning the time of crosslinking the solution after bringing it in contact with the gel. A second slow process that could be exploited in practical applications is the percolation of free polymers inside the gel. We have seen that percolation sets in as soon as the concentration of the free polymers inside the gel exceeds a threshold value which is of the order of the overlap concentration. Hence the width of the region inside the gel where free polymers percolate should be driven by diffusion and slowly
increase with time. After crosslinking, these percolating polymers could form a network that interpenetrates the gel network and strenghtens the connection between the two.

In the present work, we have studied an idealized case where the free polymers and the gel polymers were taken to have the same chemical structure. In most applications, the two will be chemically different, which will result in some incompatibility. Thus the interfaces will be more similar to regular liquid-liquid interfaces of demixed phases. The interplay of incompatibility, elasticity, and entropy in such cases will be an interesting subject of future studies.

Another aspect which should have a significant impact on the properties of the interface is chain length disparity. Here we have considered a special case where the free polymers and the strand polymers have equal lengths. In the future, it will be interesting to look at situations where the free polymers are much shorter -- leading to increased swelling of the gel -- or much longer -- leading to slower diffusion, but also, possibly, an enhanced percolation probability. Furthermore, we have considered regular networks here. Real networks are disordered and have a distribution of mesh sizes, which will also lead to interesting novel phenomena.

\section{Supporting Information}
The following files are available free of charge.
\begin{itemize}
  \item Supporting Information:  Characterization of polymers in free solution (\ref{appendix:solution}), bulk modulus of a swollen gel without polymers inside
(\ref{appendix:modulus}), additional data for isotropic gels --  monomer density profiles (\ref{SI:isodensprof}), percolation (\ref{SI:isoPerc}), motion (\ref{iso_motion}), conformation (\ref{iso_chain}) and derivation of Equation (\ref{eq:1d_diffusion}) in \ref{1d_diffusion}.
 \item Supporting Information video: Polymers diffusing into an ansiotropic regular network, clearly showing the compression and swelling of the gel. The bulk concentration of polymer solution was estimated to be $\rho_{sol_f} = 0.168 \sigma^{-3}$ .
\end{itemize}

\section{Data availability}
The codes both simulation and analysis can be accessed  at: \url{https://github.com/judevishnu/Networkdiffusion.git}.
The data that support the findings of this study will be made available upon reasonable request.

\section{Acknowledgments}
Funding from the German Science Foundation is acknowledged: SS and FS are members of the RTG 2516 (Grant No. 405552959), JV and TL are recipients of a doctoral position within the RTG 2516 program. This research was conducted using the supercomputer MOGON 2 offered by Johannes Gutenberg University Mainz (hpc.uni-mainz.de), which is a member of the AHRP (Alliance for High Performance Computing in Rhineland Palatinate,  www.ahrp.info) and the Gauss Alliance e.V.

\bigskip

\appendix
\section{Appendix: Capillary wave analysis}

\renewcommand{\thesection}{A\arabic{section}}
\renewcommand{\thesubsection}{A\arabic{section}.\arabic{subsection}}

\subsection{Determination of interfacial width}

\label{appendix:graph}

\begin{figure}[ht]
\centering
    \includegraphics[width=0.8\columnwidth]{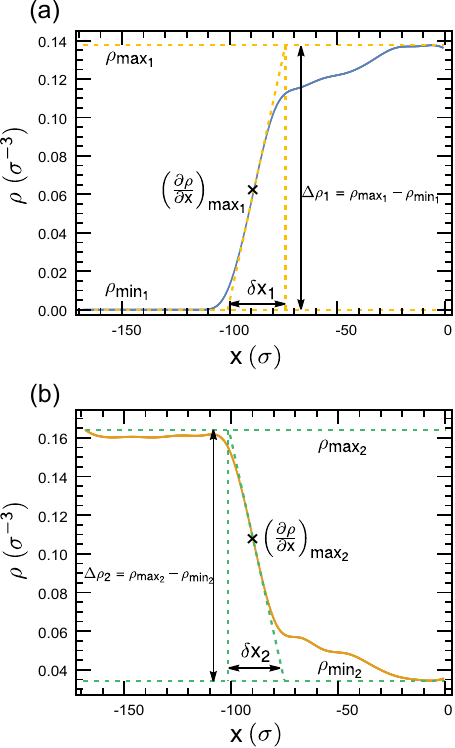}
\caption{Schematics of the method use for extracting the total interfacial width from the gel monomer density profiles (a) and from the monomer density profiles for free polymers (b). After extracting $\delta x_1$ and $\delta x_2$,  we average over them to get $w$.}
    \label{figS1:a}
    \label{figS1:b}
\label{fig:width_determination}
\end{figure}

To calculate the interfacial width in a given sub-block of the system, we employ python3  alongside with packages such as ''derivative'' \cite{kaptanoglu2022pysindy} and ''statsmodel'' \cite{seabold2010statsmodels} and proceed as follows:

\begin{itemize}
        \item First we divide the density profiles of the gel and the polymers into the left and right halves symmetrically at $x=0$. This gives four density curves.
        \item We then determine the derivative of these curves at each point (the instantaneous slope) and determine the maximum of the absolute value of this slope $\left|\frac{ \partial \rho}{ \partial x}\right|_{\text{max}}$ for each curve. In addition, we determine the maximum and minimum density values for each curve.
        \item From these numbers, we determine the interfacial width $\delta x_i$ for each
        of the four profiles via  \cite{ARDELL2012423}
        \begin{equation}
          \delta x= \frac{\rho_{\text{max}} - \rho_{\text{min}}}{\left| \left(\frac{ \partial \rho}{ \partial x} \right)_{\text{max}}  \right|}
         \end{equation}
         This definition is illustrated in Figure \ref{fig:width_determination}.
        \item Finally, we average over these four numbers to obtain the   total interfacial width
         $w =\frac{1}{4}\sum_{i=1}^{4} \delta x_i $.
\end{itemize}

\FloatBarrier

\subsection{Theoretical expression for capillary wave broadening}

\label{appendix:derivation}
Starting from the effective interface Hamiltonian, Equation \protect\eqref{eq:h}, we first perform a Fourier transform according to \protect\eqref{eq:ft}. In Fourier representation, the Hamiltonian reads

\begin{equation}\label{eq10}
    \mathcal{H}[\widetilde{h}(\vec{q})] =   \gamma A +\sum_{\vec{q}} \frac{1}{2}\Bigl(B +\gamma q^2\Bigr)\widetilde{h}(\vec{q})\widetilde{h}(-\vec{q})
\end{equation}
Now to calculate the correlations, we use the generalized equipartition theorem
 $\langle x_i \frac{ \partial \mathcal{H}}{ \partial x_j}\rangle
  = k_BT \delta_{ij}$. Here the $x_i$ are generalized coordinates.
Since the height
$h$ is a real function, it's Fourier coefficient
$\widetilde{h}(\vec{q})$ is complex conjugate to $\widetilde{h}(\vec{-q})$. Exploiting this relation yields

\begin{eqnarray}
\label{eq11}
        \langle |\widetilde{h}(\vec{q})|^2  \rangle
       & = \frac{k_BT}{(B+\gamma q^2)},
\end{eqnarray}

From \protect\eqref{eq11}, we can calculate the variance of the position of the
interface $h$ in real space as
\begin{eqnarray}
\label{eq16}
      \lefteqn{ \langle h(y,z)^2 \rangle
            =  \frac{1}{A} \: \sum_{\vec{q}}\langle \mid \widetilde{h}(\vec{q})\mid ^2\rangle }
             \qquad \nonumber\\
            &=& \frac{1}{4\pi^2}\int 2\pi q
            \: \text{d} q \:
              \langle \mid \widetilde{h}(\vec{q})\mid ^2\rangle  \nonumber\\
            &=& \frac{k_BT}{2 \pi} \int
              \text{d} q  \:\frac{q}
            {(B+\gamma q^2)} .
\end{eqnarray}
The integral in Equation \protect\eqref{eq16} can easily be solved by substitution of $p = (B+\gamma q^2)$ to get the following expression for $\langle h^2 \rangle$:

\begin{align}\label{eq17}
         \langle h^2 \rangle&=\begin{aligned}[t]\frac{k_BT}{4 \pi \gamma} \left[\ln{\Bigg(\frac{gb^2_{int}+1}{gL^2_{||}+1}\Bigg)+2\ln{\Bigg(\frac{L_{||}}{b_{int}}\Bigg)}}\right]\end{aligned}\nonumber \\
\end{align}
with $\displaystyle
   g=\frac{B}{4 \pi^2\gamma}.$
\bibliography{ms.bib}

\clearpage
\renewcommand*\contentsname{}
\twocolumn[
 \begin{@twocolumnfalse}
\begin{center}

\parbox{\textwidth}{\large \textbf{Supporting Information - Structure and dynamic evolution of interfaces between polymer solutions and gels and polymer interdiffusion: A Molecular dynamics study}}

\vspace{1cm}

\etocsettocstyle{ }{}

\localtableofcontents
\end{center}
\end{@twocolumnfalse}
]

\thispagestyle{empty}

\renewcommand{\thefigure}{S.\arabic{figure}}
\renewcommand{\thesection}{S\arabic{section}}
\renewcommand{\thesubsection}{S\arabic{subsection}}
\renewcommand{\theequation}{S\arabic{equation}}

\setcounter{figure}{0}
\setcounter{section}{0}
\setcounter{subsection}{0}
\setcounter{equation}{0}
\clearpage

\subsection{Properties of free polymer solutions }
\label{appendix:solution}

\subsubsection{Gyration radius of free chains in solution }
\label{appendix:Rg}

\begin{figure}[ht]
\centering
\includegraphics[width=0.8\columnwidth]{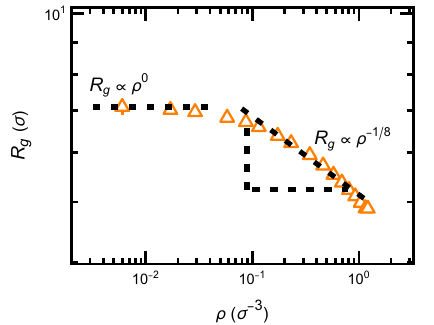}
\caption{Radius of gyration $R_g$ of free polymers in solution vs.  monomer density. Dashed lines indicates the expected power law scaling in the dilute regime ($R_g \propto \rho^0$) and in the semidilute regime ($R_g \propto \rho^{-1/8}$).
}
\label{fig:Rgvsrho}
\end{figure}

Figure \ref{fig:Rgvsrho} shows the radius of gyration of free chains ($N=102$) in solution as a function of monomer density. The theoretical value of the overlap concentration is around $\rho^* = 0.06 \sigma^{-3}$. Even though the chains are quite short, the data are consistent with a crossover from a dilute regime, where $R_g$ is independent of $\rho$, to a semidilute regime, where\cite{book_rubinstein_colby} $R_g \sim \rho^{-1/8}$.

\subsubsection[Equation of state of free solution]{Equation of state of the free solution and overlap concentration}
Figure \ref{fig:Rgvsrho} shows the pressure of the free polymer $(N=102)$ chains in a solution as a function of monomer density. We fit the data to de Gennes scaling equation\cite{book_rubinstein_colby}
\begin{equation}
\Pi \approx \Big(\frac{\rho}{N}\Big)\Big[1+\Big(\frac{\rho}{\rho^*}\Big)^{1.3}\Big],
\end{equation}
using $\rho^*$ as a fit parameter. A value of $\rho^* \approx 0.025 \sigma^{-3}$ was obtained from the fit.
\begin{figure}[ht]
\centering
\includegraphics[width=0.8\columnwidth]{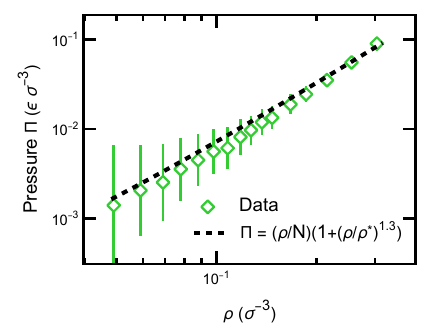}
\caption{Pressure \mbox{$\Pi$} of the free polymer solution vs. the solution density \mbox{$\rho$}. The dashed black lines shows a fit to the de Gennes scaling theory of polymers in a good solvent.
}
\label{fig:osmoticvsrho}
\end{figure}

\FloatBarrier
\subsection{Bulk modulus of bare swollen gel}
\label{appendix:modulus}
\begin{figure}[htbp]
\centering
\includegraphics[width=0.85\columnwidth]{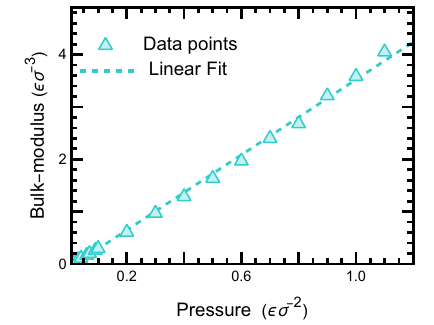}
\\
\caption{Bulk-modulus as a function of applied pressure.}
\label{fig:modulus}
\end{figure}

The elastic properties of bare gels (with no  free polymers inside) are highly
nonlinear. They are ultrasoft at small pressures and gradually stiffen at higher
pressures. Figure \ref{fig:modulus} shows the bulk modulus of a bare isotropic gel,
as determined in a fully periodic system where the gel is crosslinked across all three periodic
boundaries. The bulk modulus is calculated as
\begin{equation}
K = k_B T \frac{\langle V \rangle}{\langle V^2 \rangle -\langle V \rangle^2}
\end{equation}
The figure shows that the modulus increases almost linearly with increasing pressure.
At the pressures used to set up the gels in the simulations ($P \sim 0.001 \epsilon \sigma^{-2}$), the bulk modulus is of order $K \sim 0.003 \epsilon \sigma^{-3}$.

\subsection{Additional data for isotropic gels}
\subsubsection{Monomer density profiles}
\label{SI:isodensprof}

\begin{figure}[!t]
\centering
    \includegraphics[width=0.8\columnwidth]{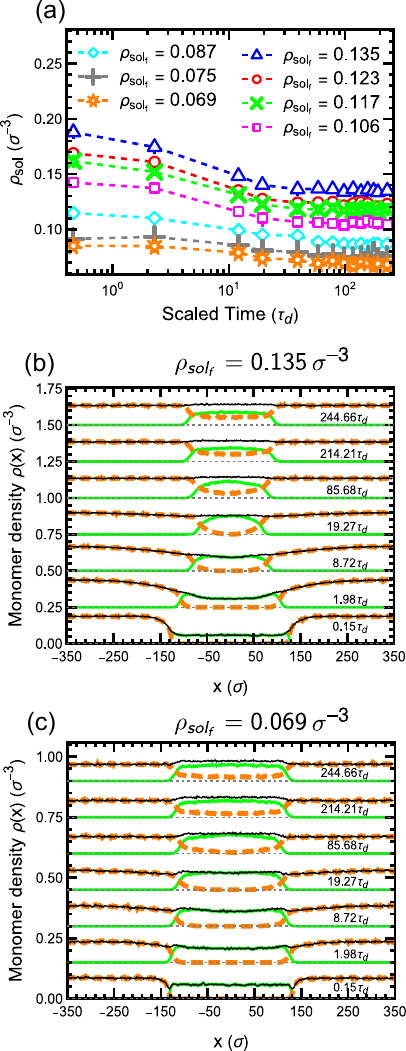}
     \caption{(a) Time evolution of monomer density in the solution phase for different
   final concentrations $\rho_{\text{sol}_f}$ (in units of $\sigma^{-3}$) as indicated.
  (b, c): Monomer density profiles across the slab for free polymers (dashed orange), gel strands (solid green), and total (solid black) at different times as indicated for different $\rho_{\text{sol}_f}$. For better readability, profiles corresponding to subsequent
  times are shifted upwards by $0.3 \sigma^{-3}$ in (b) and $0.15 \sigma^{-3}$ in (c). Thin dashed lines show corresponding baselines ($\rho=0$).
   Data correspond to isotropic gels.
  }
\label{fig:densprofiso}
\end{figure}

\begin{figure}[!t]
\centering
\includegraphics[width=0.8\columnwidth]{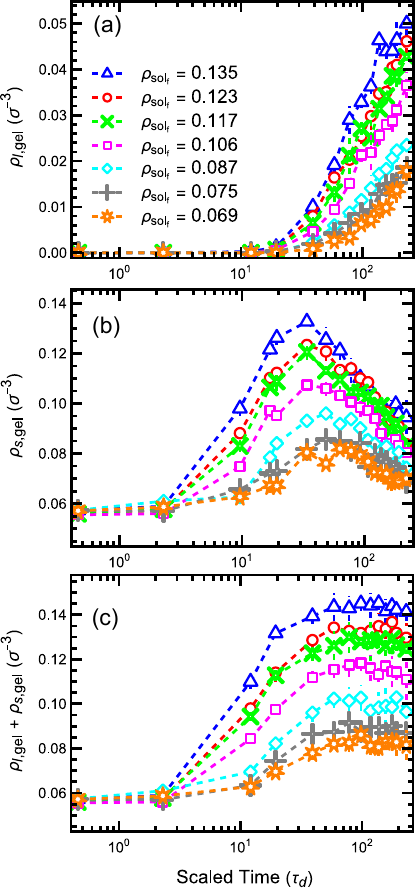}
    \caption{ Monomer densities at the center of the gel slab versus rescaled time $t/\tau_d$ for (a)  monomers belonging to free polymers, (b) gel monomers, (c) all monomers, for different final concentrations in the outer solution  $\rho_{\text{sol}_f}$ (in units of $\sigma^{-3}$) as indicated. The gel is isotropic.}
    \label{fig:figdensmida}
\end{figure}

Figure \ref{fig:densprofiso} shows complementary data to Figure \ref{fig:densprofaniso} in the main text:
The time-dependent reduction of polymer concentration in the solution as the polymers diffuse into the
gel (Figure \ref{fig:densprofiso} (a) and two selected sets of monomer density profiles across the slab, one at low concentration and one at high concentration. Figure \mbox{\ref{fig:figdensmida}} (a)- (c) is complementary to Figure \mbox{\ref{fig:figdensmidaniso}} (a)-(c) in the main text and shows the density of free polymers, the gel strand density and the total monomer density in the middle of the slab as a function of time in units of the chain relaxation time $\tau_d$. The figure shows that the system as a whole does not reach equilibrium during the simulation time, as in the case of the anisotropic gel.

\subsubsection{Percolation}
\label{SI:isoPerc}
\begin{figure}[!t]
    \includegraphics[width=0.8\columnwidth]{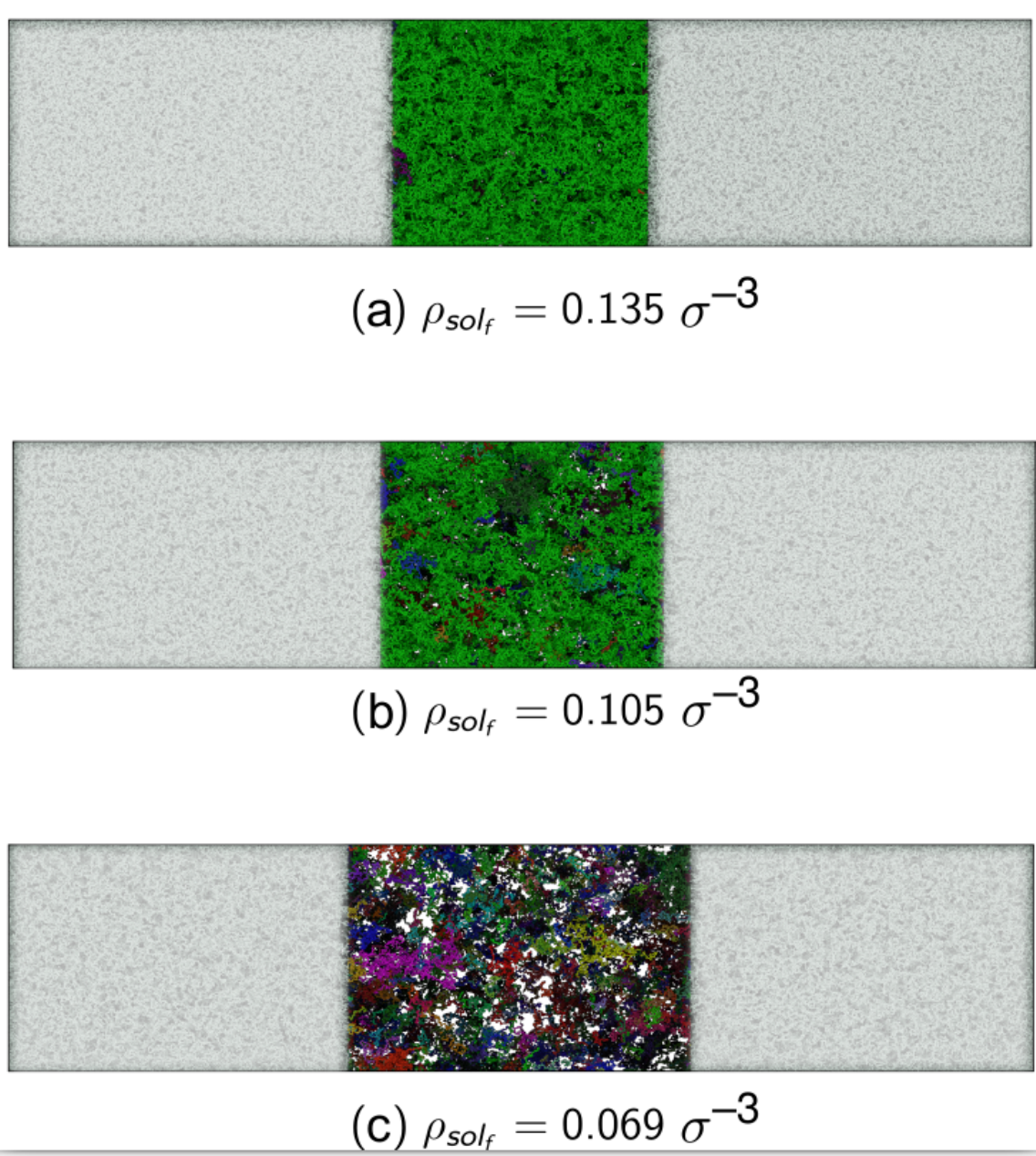}
    \caption{(a-c) Snapshots showing cluster formation of free chains within the gel region as viewed from top of the box (along z-axis) at different final polymer densities $(\rho_{sol_f})$ as indicated. Different clusters are colored differently in order to be distinguishable. The light gray color marks the solution region. The gel is isotropic. }
   \label{fig:percisosnapshots}
\end{figure}

\begin{figure}[tbhp]
    \includegraphics[width=0.8\columnwidth]{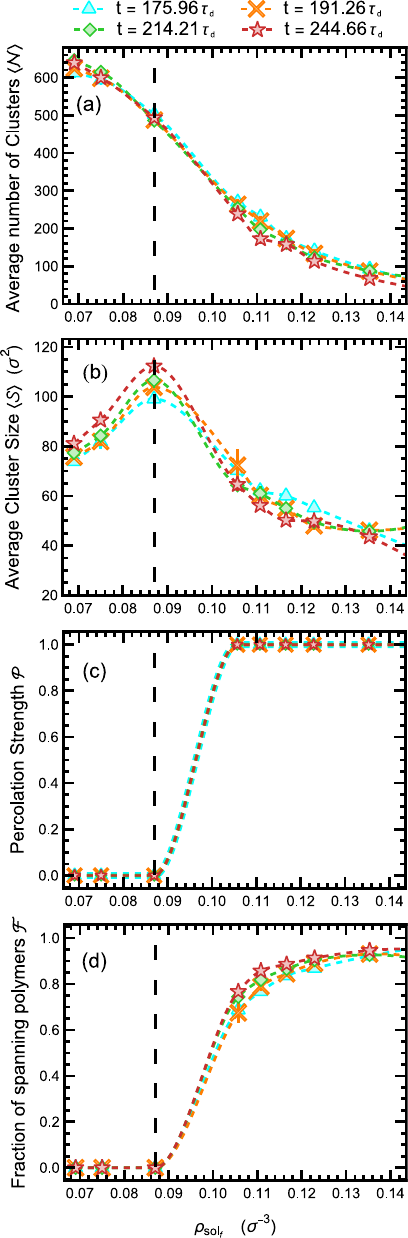}
    \caption{ Characteristics of cluster distributions vs.\ $(\rho_{sol_f})$ at different times as indicated: (a) Average number of clusters; (b) Average cluster size (mean squared radius of gyration) of finite clusters; (c) Percolation strength (see text); (d) Fraction of free polymers inside gel that are part of the spanning cluster. The gel is isotropic.}
   \label{fig:perciso}
\end{figure}

Figures \ref{fig:percisosnapshots} and \ref{fig:perciso} show complementary snapshots and data to Figure \ref{fig:percAnsio} in the main text to characterize the cluster formation and the percolation of free chains inside the isotropic gel.

\subsubsection{Monomer motion profiles}
\label{iso_motion}

\begin{figure}[htbp]
\centering
    \includegraphics[width=0.85\columnwidth]{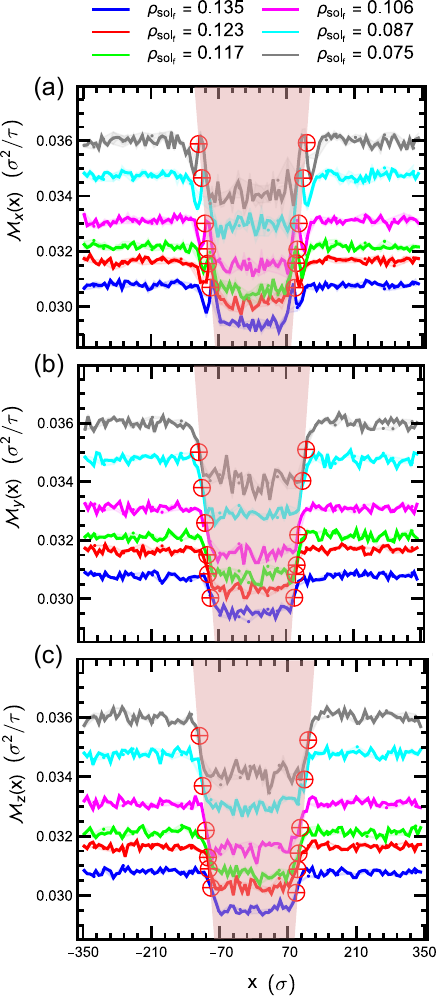}
\caption{ Profiles of monomer motion  $\mathcal{M}_{_\alpha}(x)$  in $\alpha=x$ (a), $\alpha=y$ (b) and $\alpha=z$ (c) direction vs. $x$-coordinate of monomers. The red symbols $\bigoplus$ indicate the approximate position of interface, the red shading, the gel regions. The gel is isotropic.}
\label{fig:isodiff}
\end{figure}

Figure \ref{fig:isodiff} shows complementary data for the isotropic gel to the profiles of local monomer motion in the anisotropic gel in Figure \ref{fig:anisodiff} in the main text. Different from Figure \ref{fig:anisodiff}, the profiles of ${\cal M}_z(x)$ and ${\cal M}_y(x)$ are the same. Inside the gel slab, the monomer motion is the same within the error in all three directions.

\subsubsection{Chain conformation profiles}
\label{iso_chain}
\begin{figure}[htpb]
\centering
    \includegraphics[width=0.8\columnwidth]{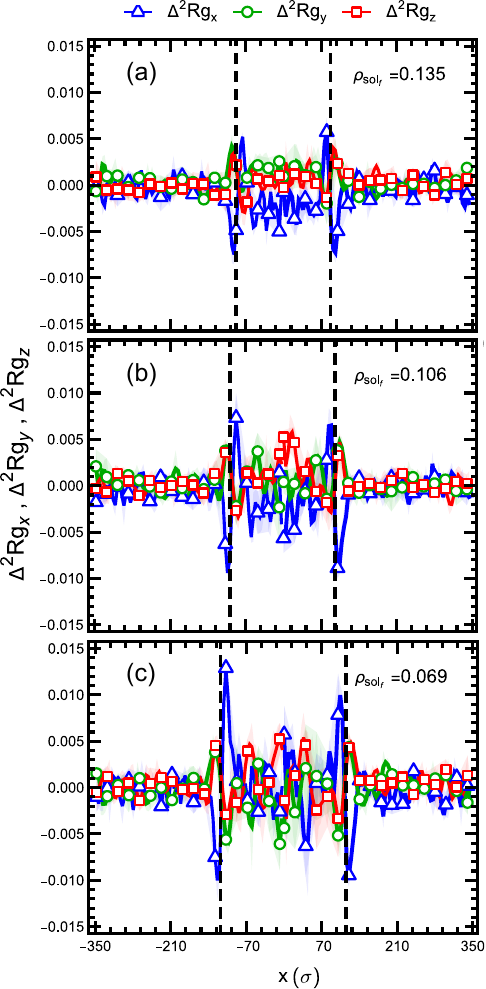}
  \caption{Profiles of chain conformation parameters $\Delta^2 {R_g}_{_\alpha}$ (see text) across the gel slab for different concentrations  $\rho_{\text{sol}_{f}}$ as indicated. The $x$ coordinate refers to the center of mass of chains. Black dashed line indicates the position of the interface, shaded bands the error.  The gel is isotropic.}
\label{fig:isoconf}
\end{figure}

Figure \ref{fig:isoconf} shows complementary data for the isotropic gel corresponding to the profiles of chain conformation parameters $\Delta^2 {R_g}_{x,y,z}$ in anisotropic gels shown in Figure \ref{fig:anisoconf} in the main text. Different from the anisotropic gel, the chain conformations are isotropic within the error inside the gel. At the surface, a small change in chain orientation is observed for  $\Delta^2 {R_g}_x$ as seen in the anisotropic gels.

\FloatBarrier

\subsection{Solution of diffusion equation for free polymers diffusing into a gel slab}
\label{1d_diffusion}
\begin{figure}[h!]
\centering
    \includegraphics[width=0.67\columnwidth]{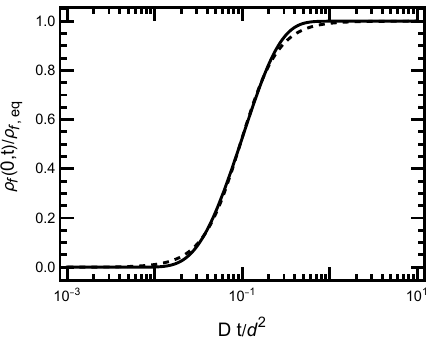}
  \caption{Comparison of equation (\protect\ref{eq:rhof_1d_center}),
   describing the density of diffusing monomers at the center
   of the slab as a function of time (solid line), with the approximate expression
   Equation (\protect\ref{eq:rhof_1d_center_approximate}) (dashed line).}
   \label{fig:1dDiffeq}
\end{figure}

To estimate the time scales required for the full equilibration of our system, we assume that the longest time scale is set by the diffusion of free polymers into the gel. We consider a system of polymers diffusing into a gel of thickness $d$ from both sides with diffusion constant $D$. The gel is centered at $x=0$, and the polymer density at the surface of the slab is set at $\rho_f(\pm d/2,t) \equiv \rho_{f,\text{eq}}$ (Dirichlet boundary conditions), where $\rho_{f,\text{eq}}$ is the equilibrium density of free polymers in the slab after equilibration. The solution of the one-dimensional diffusion equation ($\partial_t \rho = D \: \partial_{xx} \rho$) with these boundary condition at $x = \pm d/2$ and initial condition $\rho_f(x,t=0)\equiv 0$ is
\begin{eqnarray}
\label{eq:rhof_1d}
    \rho_f(x,t) &=& \rho_{f,\text{eq}}\bigg[ 1
  -\:  \frac{4}{\pi}
    \sum_{n=0}^\infty \frac{(-1)^{n}}{2 n+1}
      \\ && \nonumber
      \times \cos\big( \frac{\pi}{d} (2n+1) x \big)
\\ && \nonumber
   \times  \exp\Big(
      - D \: (\frac{\pi}{d})^2  (2 n + 1)^2\: t \Big)
    \bigg].
\end{eqnarray}
 This results in the time evolution equation
 \begin{equation}
 \label{eq:rhof_1d_center}
\frac{\rho_f(0,t)}{\rho_{f,\text{eq}}} = \bigg[
 1 - \varphi\Big((\frac{\pi}{d})^2 D \:t\Big) \bigg]
 \end{equation}
\begin{displaymath}
   \mbox{with} \quad \varphi(\tau) = \frac{4}{\pi}
    \sum_{n=0}^\infty \frac{(-1)^{n}}{2 n+1}
    \:  \textrm{e}^{ - \tau  (2 n + 1)^2}.
\end{displaymath}
for $\rho_f(x,t)$ at the center of the slab.
The function $\varphi(\tau)$ is initially close to zero up to
$\tau \sim 0.1$, then it increases quasi logarithmically,
until it finally saturates around $\tau \sim 5$. The time
range between the initial discernible increase of free
polymer density in the center of the gel, and the final saturation
thus spans roughly 1.5 orders of magnitude. Numerically, the
function $\rho_f(0,t)$ is reasonably well approximated
by
\begin{equation}
\label{eq:rhof_1d_center_approximate}
   \frac{\rho_f(0,t)}{\rho_{f,\text{eq}}}  \approx
   \: \frac{1}{2} \Big[
     \tanh \Big( \ln(\frac{Dt}{d^2}) + 2.357 \Big) + 1 \Big],
\end{equation}
as shown in Figure \ref{fig:1dDiffeq}.

\end{document}